\documentclass{aa}  

\usepackage{graphicx}
\usepackage{txfonts}
\usepackage[colorlinks=true,
     linkcolor=blue,
     filecolor=blue,
     citecolor = blue,      
     urlcolor=blue,]{hyperref}
\usepackage{graphicx}	
\usepackage{amsmath}	
\usepackage{amssymb}	
\usepackage{siunitx}
\usepackage[dvipsnames]{xcolor}
\usepackage{float}

\addto\extrasenglish{}
\addto\extrasenglish{}
\addto\extrasenglish{}
\addto\extrasenglish{}
\addto\extrasenglish{}
\addto\extrasenglish{}
\addto\extrasenglish{}
\addto\extrasenglish{}
\newcommand{\aref}[1]{\hyperref[#1]{Appendix~\ref{#1}}}

\definecolor{darkgreen}{rgb}{0.13, 0.55, 0.13}

\makeatletter
\renewcommand*\aa@pageof{, page \thepage{} of \pageref*{LastPage}}
\makeatother

\begin{document}

   \title{The role of magnetic fields in disc galaxies: spiral arm instability}


   \author{Raghav Arora
          \inst{1},
          Christoph, Federrath \inst{2}\fnmsep, 
          Robi Banerjee \inst{1} , 
          \and 
          Bastian K{\"o}rgten \inst{1} ,
          }

   \institute{Hamburger Sternwarte, Universit\"at Hamburg, Gojenbergsweg 112, 21029 Hamburg, Germany\\
              \email{raghav.arora@hs.uni-hamburg.de}
         \and
             Research School of Astronomy and Astrophysics, The Australian National University, Canberra, ACT 2611, Australia\\
             }

   \date{Received xxxx; accepted xxxx}
   
\titlerunning{Magnetised spiral arm instability}
\authorrunning{Arora et al.}

 
  \abstract
   {Regularly-spaced, star-forming regions along the spiral arms of nearby galaxies provide insight into the early stages and initial conditions of star formation. The regular separation of these star-forming regions suggests spiral arm instability as their origin.}
   {We explore the effects of magnetic fields on the spiral arm instability.}
   { We use three-dimensional global magnetohydrodynamical simulations of isolated spiral galaxies, comparing three different initial plasma $\beta$ values (ratios of thermal to magnetic pressure) of $\beta=\infty$, $50$, and $10$. We perform Fourier analysis to calculate the separation of the over-dense regions formed as a result of the spiral instability. We then compare the separations with observations.}
   {We find that the spiral arms in the hydro case ($\beta = \infty$) are unstable, with the fragments initially connected by gas streams, reminiscent of Kelvin-Helmholtz instability. In the $\beta = 50$ case, the spiral arms also fragment, but the fragments separate earlier and tend to be slightly elongated in the direction perpendicular to the spiral arms. However, in the $\beta = 10$ run the arms are stabilised against fragmentation by magnetic pressure. Despite the difference in the initial magnetic field strengths of the $\beta = 50$ and $10$ runs, the magnetic field is amplified to $\beta_\mathrm{arm} \sim 1$ inside the spiral arms for both runs. The spiral arms in the unstable cases (hydro and $\beta=50$) fragment into regularly-spaced, over-dense regions. We determine their separation to be $\sim 0.5$~kpc in the hydro and $\sim 0.65$~kpc in the $\beta = 50$ case, both in agreement with the observed values found in nearby galaxies. We find a smaller median characteristic wavelength of the over-densities along the spiral arms of $0.73^{+0.31}_{-0.36}$~kpc in the hydro case, compared to $0.98^{+0.49}_{-0.46}$~kpc in the $\beta = 50$ case. Moreover, we find a higher growth rate of the over-densities in the $\beta = 50$ run compared to the hydro run. We observe magnetic hills and valleys along the fragmented arms in the $\beta = 50$ run, which is characteristic of the Parker instability. }
   {}


   \maketitle
%

\section{Introduction}

Star formation in spiral galaxies is driven by their spiral arms \citep{ngan_pettitt_2017}. Star-forming regions that resemble beads-on-a-string pattern are ubiquitous in spiral arms of numerous nearby galaxies, despite having a broad range of morphological features and physical properties \citep[e.g.][]{elmegreen_1983, elmegreen_2006}. Recently, also seen in rings of barred \citep{gusev_shimanocskayay_2020} and lenticular galaxies \citep{proshina_moiseev_silchenko_2022}. These star-forming regions are found to be bright in IR \citep{elmegreen_2006, elmegreen_2018, elmegreen_2019}, and at times also in FUV and $\rm H\alpha$ \citep{efremov_2009, efremov_2010, gusev_efremov_2013, gusev_2022} emission,  tracing various stages of star formation. Despite the variety in spiral galaxies that host these regular star-forming regions, the adjacent separations of these regions fall in the range $350-500~\si{pc}$ and (or) integer multiples of this range \citep[see][]{gusev_2022}. For example, it was found that adjacent separation of star-forming regions in the spiral arms of NGC 628 and M100 were both $\sim 400\, \si{pc}$ \citep{gusev_efremov_2013, elmegreen_2018}. Another prominent example being the North-Western arm of M31, where star complexes with a spacing twice this value $\sim 1.1\, \si{kpc}$ were found \citep{efremov_2009, efremov_2010}. The presence of this characteristic separation of these star-forming regions in spiral arms hints that their origin could be through a spiral arm instability. Another interesting feature observed in unison with regular star-forming regions in M31 were the presence of magnetic fields out of the plane of the galaxy with a wavelength twice that of the adjacent separation of these regions \citep{efremov_2009, efremov_2010}. This indicates that magnetic fields could play an important role in their formation. 

Beads-on-a-string patterns have been observed along spiral arms in numerical simulations \citep[e.g.][]{chakrabarti_2003, wada_koda_2004, kim_ostriker_2006, shetty_ostriker_2006, wada_2008, renaud_2013}. The physical mechanism responsible, however, has been debated. It was hypothesized to be the Kelvin-Helmholtz (KH) hydrodynamical instability \citep{wada_koda_2004}, an artifact of numerical noise \citep{hanawa_kikuchi_2012} or of infinitesimally flat 2D discs \citep{kim_ostriker_2006}, a magnetic-jeans instability (MJI) \citep{kim_ostriker_2006} which was also called the feathering instability \citep{lee_kit_2014}, or a hydro instability along spiral shocks distinct from the KH instability named as the Wiggle Instability \citep[WI]{kim_wong_kim_2014}. Its physical nature was established recently to be a combination of KHI \citep[as proposed in][]{wada_koda_2004} and a vorticity generating instability due to repeated passages of spiral shocks \citep{mandowara_sormani_2021}. Using a subset or restricting to different physical processes, these works saw facets of the same physical instability around the spiral shock. However, we also know from 2D local simulations and linear analysis \citep{mandowara_sormani_2021}, that the spiral instability is highly non-linear and sensitive to the physical properties of the interstellar medium (ISM). This includes the sound speed ($c_{s}$), the strength of the spiral shock etc. One such important factor are the magnetic fields.

Even though the role of magnetic fields has been explored, it still remains to be seen what effect they can have in a global realistic setting. We expect from earlier works that they can affect both the length scales and the growth scales of the instability. For example, it was shown in \cite{lee_kit_2014} that the spiral shock fronts were weaker in the presence of stronger magnetic fields, not affecting the feather instability's growth rate. However, the range of their parameter space spanned only from $\beta = 2-4$.
In \cite{kim_wong_kim_2014}, it was found in 2D local linear analysis and simulations that equipartition magnetic fields ($\beta = 1$) in comparison with very weak fields ($\beta = 100$) decreased the growth rate of the spiral arm instability by a factor $4$ and increased the wavelength of the dominant mode by a factor of $2$. Thus having an overall stabalising effect. In three-dimensional local box simulations, \cite{kim_ostriker_2006} found no vorticity generating instability and attributed it to presence of magnetic fields ($\beta = 1, 10$), but found that the spiral shocks were unstable nonetheless in presence of self-gravity and magnetic fields. This was also confirmed in global 2D simulations in \cite{shetty_ostriker_2006} where equipartition magnetic field strengths prevented the pure hydro vortical instability, but the spiral arms fragmented in presence of self-gravity despite the magnetic fields. Most of these studies are 2D \citep{lee_kit_2014, kim_wong_kim_2015, shetty_ostriker_2006} or local \citep{lee_kit_2014, kim_wong_kim_2015, kim_ostriker_2006} and all of them are isothermal. While there are more sophisticated magnetised galaxy simulations \citep{dobbs_price_2008, khoperskov_2017}, they have focused on the global evolution of the disc, rather than on the spiral arm instability. It is yet to be seen how magnetic fields affect the spiral arm instability in 3D global disc galaxy simulations where one resolves the spiral arm instability, MJI, and also the Parker instability \citep{parker_1966} that arises due to the stratified nature of the medium perpendicular to the disc. 

In this study, we focus on the physical effects of magnetic fields on the spiral arm instability and its impact on the spacings of the over-densities that result in the spiral arms. We perform three-dimensional self-gravitating magnetised disc galaxy simulations that employ fitted-functions for equilibrium cooling, heating and an external spiral potential. This allows us to capture the spiral arm instability in a more realistic environment and study the effects of magnetic fields on it. For this purpose, we build a library of simulations with varying initial magnetisation.

The paper is organized as follows: \autoref{sec:method} describes our library of simulations. In \autoref{sec:results},  we compare the the models with different magnetisation and focus on the basic morphology of our galaxies, the approximate timescales of the spiral arm instability and the cloud spacings of the unstable spiral arms. We discuss the caveats of this work, the role played by magnetic fields on the spiral arm instability and compare them with existing observations and simulations in \autoref{sec:Discussion}. We summarise our results in \autoref{sec:conclusions}.

\section{Methods} \label{sec:method}

\subsection{Simulation Setup} \label{sec:simulationSetup}

Our 3D MHD disc galaxy simulations have self-gravity, an external spiral potential, magnetic fields and fitted functions for optically thin cooling and heating that include heating from cosmic and soft X-rays, the photoelectric effect, as well as the formation and dissociation of $\rm H_{2}$ \citep[see][]{bastian_2019}. The cooling and heating gives us a multiphase ISM
consisting of the warm neutral medium (WNM), cold neutral medium (CNM), and a cold molecular medium (CMM). These form self-consistently in our simulations. The cooling and heating curve also has the thermally unstable regime, which is important for molecular cloud formation, in the density range $1 \leq n / \rm cm^{-3} \leq 10$. Our minimalist global models strike a balance by including the dominant physical effects such as self-gravity, galactic shear, thermal instabilities and magnetic fields, while at the same time avoiding complicated stochastic effects such as star formation and various feedback mechanisms. We do not study these mechanisms here since we want to focus on the effects of magnetic fields and defer an investigation of the effects of feedback to a future study.   

The system of equations that we solve are  - 
\begin{align}
    \frac{\partial \rho}{\partial t} + \nabla.(\rho \textbf{v}) &= 0, \\
    \frac{\partial (\rho \textbf{v})}{\partial t} + \nabla. (\rho \textbf{v} \textbf{v}) = -\nabla P - \nabla \left (\Phi_{\rm ext}+ \Phi \right ) & + \frac{1}{4\pi} (\nabla \times \textbf{B})\times\textbf{B}, \\ 
    \frac{\partial \textbf{B}}{\partial t} &= \nabla \times \left ( \textbf{v} \times \textbf{B} \right ), \\ 
    \nabla^{2} \Phi &= 4\pi G \rho, \\
    \begin{split}
    \frac{\partial }{\partial t} \left ( \frac{\rho v^{2}}{2} + \rho \epsilon_{\rm int} + \frac{B^{2}}{8\pi} \right ) +  &\\
    \nabla. \left [ \left ( \frac{\rho v^{2}}{2} + P + \rho \epsilon_{\rm int} + \frac{B^{2}}{8\pi}\right ).\textbf{v} + v_{j}\mathcal{M}_{ij}\right] & = \frac{\rho}{m_{\rm H}} \Gamma  - \left ( \frac{\rho}{m_{\rm H}} \right )^{2} \Lambda(T),
    \end{split}
\end{align}

where $\rho$ is the density and $\textbf{v}$, $P$, $\textbf{B}$ are the velocity, pressure and the magnetic field of the gas. The gravitational potential of the gas and the external gravitational potential are given by $\Phi, \Phi_{\rm ext}$. $m_{\rm H}$ is the mass of Hydrogen atom and $\Gamma$, $\Lambda$ are the heating and the cooling rates respectively \citep{koyama_inutsuka_2002, VazquezSemadeniEtAl2007}. $\mathcal{M}_{ij} = \frac{B^{2}}{8\pi} \delta_{ij} - \frac{B_{i}B_{j}}{4\pi}$, and we use the 
polytropic equation of state, which gives $\epsilon_{\rm int} = P/\rho (\gamma - 1)$, where $\gamma = 5/3$~. These represent the ideal MHD equations where we have neglected the magnetic diffusivity and fluid viscosity. 

We use the \textsc{flash} \citep{FryxellEtAl2000,dubey_fisher_2008} grid-based magnetohydrodynamical code for performing the simulations. The disc is initialized at the centre of a cuboidal box with side length $L_{xy} = 30$~kpc in the plane of the disc and $L_{z} = 3.75$~kpc in the direction perpendicular to it. The minimum cell size of the base grid is $ 234$~pc, i.e., the base grid has a resolution of $128\times128\times16$ cells. However, we achieve a maximum resolution corresponding to a minimum cell size of $ 7.3$~pc by using adaptive mesh refinement (AMR) with 5 levels of refinement corresponding to a maximum effective resolution of $4096\times4096\times512$ cells, such that the local Jeans length is resolved with at least 32~grid cells \citep{FederrathEtAl2011} for $R \geq 5$~kpc. We do this in two steps. First we use 4~levels of refinement for the initial $0.3$ $T_{\rm rot}$ ($100$~Myr) of the evolution. We then increase the maximum refinement level to 5, which saves computational costs in the initial phases of the evolution, and allows us to achieve an overall higher refinement when the spiral arms start to develop. The initial Jeans length in our models is $\sim 1.8\,\si{kpc}$, and thus we resolve it by 2 additional levels of refinement at the start. In order to avoid artificial fragmentation on the highest level of refinement due to violation of the Truelove criteria \citep{TrueloveEtAl1997}, we have an artificial pressure term that is adjusted so that the local jeans length is resolved with at least four grid cells \citep{bastian_2019}.   

\subsubsection{Basic setup}

As done in earlier studies \citep[see][]{bastian_2018, bastian_2019}, we initialize the disc galaxies in our simulation suite by keeping the effective Toomre parameter ($Q_{\rm eff}$) constant, defined as 
\begin{equation}\label{eqn:q_eff}
        Q_{\rm eff} = \frac{\kappa \left( c_{\rm s}^{2} + v_{\rm a}^2 \right )^{1/2}}{\pi G \Sigma},
\end{equation}
where $\kappa = \sqrt{2}v_{\rm c}/r$ and $\Sigma$ are the epicyclic frequency and the surface density of the disc, $c_{\rm s}$ is the sound speed, and $v_{\rm a}$ is the Alfv\'en speed of the medium. We do this so that all the simulations have a similar response to axisymmetric gravitational perturbations. Next, we use the scale height radial profile, $H(R)$, from $\rm H\, \textsc{i}$ observations of the Milky Way \citep{binney_1998}, which gives us 
\begin{equation} \label{eqn:density}
\rho (R, z) = \frac{\kappa c_{\rm s} \sqrt{1 + 2\beta^{-1}}}{ 2 \pi G Q_{\rm eff} H(R)} \mathrm{sech^2}\left (\frac{ z}{ H(R)}\right ) 
\end{equation}
where $H(R) = R_{\odot}(0.0085 + 0.01719 R/R_{\odot} + 0.00564 (R/R_{\odot})^{2} )$ with $R_{\odot} = 8.5 \rm ~kpc$ and $\beta = 2c_{\rm s}^{2}/v_{\rm a}^{2}$ is the plasma-beta (ratio of thermal to magnetic pressure) of the disc. For numerical reasons, we define the inner $5$~kpc region to be gravitationally stable and keep it unresolved with $Q_{\rm eff} = 20$. We focus on the disc with $R>5$~kpc, where we have the initial $Q_{\rm eff} = 3$, making the region of interest gravitationally stable to axisymmetric perturbations since $Q\geq1$. We also have pressure equilibrium at the boundaries to avoid any gas inflows and outflows due to the same. We further apply a buffer zone of $1\,\si{kpc}$ from the inner disc for our analysis to avoid any boundary effects.

We adopt a flat rotation curve for our galaxies, with the circular velocity in the plane of the galaxy given by 
\begin{equation}\label{equation:rotationCurve}
     v_{\rm rot} = v_{\rm c}\frac{R}{\sqrt{R_{\rm c}^{2} + R^{2}}}, 
\end{equation}
where $v_{\rm c} = 200 \rm ~ km \, s^{-1}$ for Milky-Way-like galaxies and $R_{\rm c} = 0.5 ~\rm kpc$ is the core radius. This is the exact solution for the adopted dark matter potential
\begin{equation} \label{eqn:darkMatter_pot}
    \Phi_{\rm dm} = \frac{1}{2}v_{\rm c}^{2} \ln{ \left \{ \frac{1}{R_{\rm c}^{2}} \left [ R_{\rm c}^{2} + R^{2} + \left (  \frac{z}{q} \right )^2 \right ] \right \} }.
\end{equation}
We use a marginally lower value of $v_{\rm c} = 150~\si{km\:s^{-1}}$ compared to the Milky Way. We do this to isolate and focus on the effects of the spiral instability from the presence of swing instabilities in a low-shear environment.

\subsubsection{Turbulent initial conditions}
In addition to the circular velocity of the disc, we add a turbulent initial velocity field with $v_{\rm rms} = 10 \rm ~ km \, s^{-1}$ and a Kolmogorov scaling of $k^{-5/3}$ on scales $\left [ 50, 200\right ] $~pc. The turbulent velocity field was constructed to have a natural mixture of solenoidal and compressible modes, generated with the methods described in \citet{FederrathDuvalKlessenSchmidtMacLow2010}, using the publicly available \texttt{TurbGen} code \citep{FederrathEtAl2022ascl}. The details of the initial turbulent perturbations are not critical for our numerical experiments and they primarily serve to break the symmetries in the idealised setup. They also ensure that the perturbations for the instabilities under investigation here to develop self-consistently rather than being seeded by numerical noise. After the initial turbulent seeds have decayed, ISM turbulence is subsequently driven primarily by gas flows and spiral arm dynamics.

\subsubsection{Magnetic field}
Magnetic fields are observed in nearby disc galaxies to be roughly in equipartition with the turbulent kinetic energy and in super-equipartition with the thermal energy in the ISM \citep[and references therein]{beck_2015}. We characterise the strength of the magnetic field with the plasma-beta ($\beta$), that is, the ratio of the thermal to the magnetic pressure of the medium, which is also equal to the ratio of the thermal to magnetic energies. Our initial values are $\beta \in \left \{\infty, 50, 10 \right \}$, which represent the hydro, weak and moderate magnetisation cases of our disc. We choose a higher value of $\beta$ than observed ($\beta_{\rm obs} \leq 1$), because we expect it to decrease with the dynamical evolution of the galaxy. We initialise the magnetic fields to be completely toroidal ($m = 0$ mode), which is the dominant mode found in galaxies \citep{beck_chamandy_elson_2020}, with a dependence on the gas density, such that $B\propto n^{\alpha}$, where $\alpha = 0.5$ \citep[as done in][]{bastian_2019}.

\subsubsection{Spiral potential}
For generating the spiral arms, we adopt a rigidly rotating two-armed spiral potential \citep{cox_gomez_2002} with a pattern speed of $13.34$~\si{km. .s^{-1}.kpc^{-1}}, which gives us a co-rotation radius of $= 11.25$~\si{kpc}, and a pitch angle of $\alpha = 20 \si{\degree}$. Thus, the external gravitational potential can be written as 
\begin{equation}
    \Phi_{\rm ext} = \Phi_{\rm dm} + \Phi_{\rm  sp},
\end{equation}
where $\Phi_{\rm dm}$ is the dark matter potential that provides the flat-rotation curve (\autoref{eqn:darkMatter_pot}) and $\Phi_{\rm sp}$ is the spiral potential \citep[see eq.~8 in][]{cox_gomez_2002}. We chose the amplitude of the spiral such that, on average, the magnitude of the force due to the potential is $\sim0.4$ times that of the dark matter potential, analytically given by
\begin{equation} \label{eqn:f_sp_const}
  \mathcal{F_{\rm sp}} = \left \langle \frac{\langle f_{\rm sp} \rangle_{\rm \phi}}{f_{\rm dm}} \right \rangle_{\rm r} \sim 0.40,  
\end{equation}
where $\langle f_{\rm sp}\rangle_{\rm \phi}$ is the azimuthal average of $f_{\rm sp} = \nabla \Phi_{\rm sp}$ and $\langle\dots\rangle_{r}$ denotes the radial average, taken over $6$ to $11\,\si{kpc}$, which is our region of interest. 

\subsubsection{Simulation parameter study}

Our library of simulations contains three runs having different initial magnetisation, with $\beta \in \left \{\infty, 50, 10 \right \}$, all with the same strength of the spiral arm perturbation, $\mathcal{F}_{\rm sp} = 0.4$, rotational velocity of $v_{\rm c} = 150\,\si{km\,s^{-1}}$, and an effective $Q = 3$. From \autoref{eqn:density}, it follows that fixing these values leads to an initial density field that differs only slightly (by $\leq 2 \%$) between the simulations with different $\beta$. The key initial parameters are summarised in \autoref{table:init}, where we also show the mass-weighted average density ($n$), the sound speed ($c_{\rm s}$) and the magnetic field magnitude of the disc region of interest. Note that with this parameter set, we initialise our galaxies to be marginally stable to axisymmetric gravitational instabilities, the thermal instability as well as swing instabilities. Since we have an initial $Q> 1$, mid-plane density of a factor of two less than the thermally unstable regime, and a low-shear environment with $v_{c} = 150 \, \si{km\, s^{-1}}$.

\begin{table*}
\caption{Initial conditions of the simulations. The $c_{\rm s}$ and the $B_{\rm mag}$ are the mass-weighted averages of the disc region between $r = 6\,\si{kpc}$ and $10\,\si{kpc}$ and $ |z| \leq H (\text{at}\:R = 8.5 \,\si{kpc})$, calculated at $t=2 \, \si{Myr}$.}
\label{table:init}
\begin{tabular}{ccccccccc}
\hline
\hline
Simulation Number & $\beta$ & $Q$  & $n$ $(\si{cm^{-3}}  )$ & $v_{\rm c} \,( \rm km \, s^{-1})$ &  $c_{\rm s}\, (\rm km \, s^{-1})$ & $\textbf{B}$ ($\mu$G) & $\mathcal{F_{\rm sp}}$ & Spiral arm stability\\
\hline
\hline
      1 &  $\infty$ &  3 & 0.58 & 150 & 8.13 & 0 & 40\% & $\rm Unstable$ \\
      2    &  50   &  3 & 0.59 &  150 &  8.11 &  0.69 & 40\% & $\rm Unstable$ \\
      3 & 10 & 3 & 0.64 & 150 &  8.01 & 1.60 & 40\% & $\rm Stable$ \\
\hline
\hline
\end{tabular}
\end{table*}

\section{Results} \label{sec:results}

Here, we discuss the main results of our simulations. First we show the basic evolution of the three runs, that is, their general morphology and time scales of spiral arm formation and fragmentation in \autoref{subsect:basicsEvolution}. In \autoref{result_subsec:cloud_spacings_and_fragmentation_modes}, we focus on the cases where the spiral arms fragment into over-dense regions. Here, we also describe our method for extracting the separation of the clouds and then report them as well. We then showcase the physical properties of the spiral arms and the effects that magnetic fields have on them in \autoref{result_subsect_spiral_arm_properties}.

\begin{figure*}
    \centering
    \includegraphics[width=0.99\linewidth]{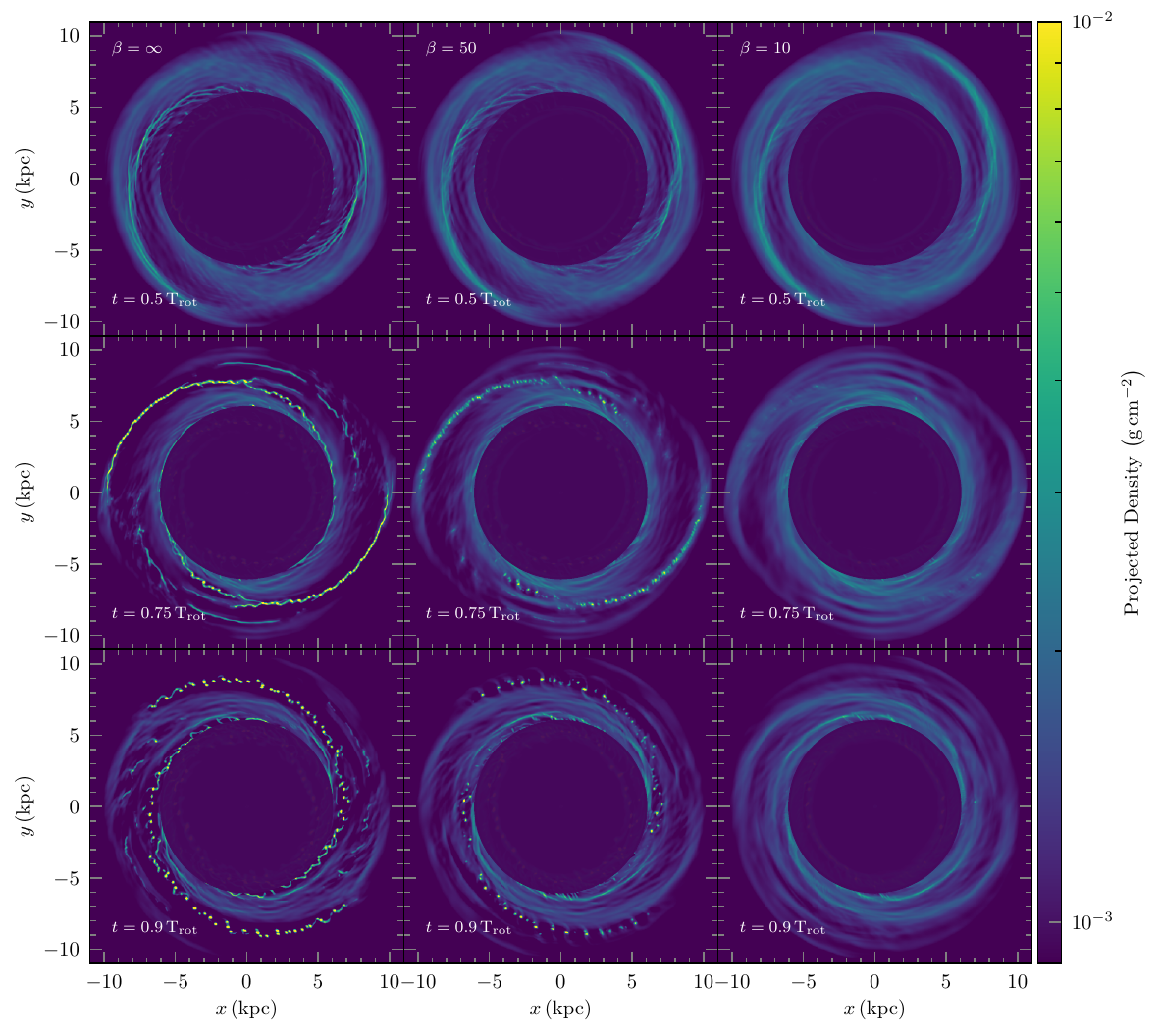}
    \caption{Each panel shows the column density of the disc projected onto the z-plane. The columns represent the initial plasma $\beta$ (ratio of thermal to magnetic pressure) of the runs and the rows from top to bottom show the evolution of the galaxy. Roughly depicting the times when the spiral arms first form, when the arms are fragmenting into over-densities and when they separate out into clouds. We can see that the disc gets more diffuse with decreasing plasma $\beta$ in the first row. The second and the third row highlights the effects of magnetic fields on the stability of the spiral arms, where we see that the $\beta = 10$ run is stable to fragmentation, while the hydro and $\beta = 50$ cases fragment and yet still have noticeable morphological differences. 
    }
    \label{fig:surfaceDensity_pBeta}
\end{figure*}

\subsection{Basic Evolution} \label{subsect:basicsEvolution}
Here we describe the basic morphology and evolution of our simulations. We first discuss the spiral arms that form self-consistently, and then their subsequent fragmentation patterns in \autoref{subsubsect: morphologicalEvolution}. We then quantify the timescales over which we see this evolution in \autoref{results_subsect:timeEvol}.

\subsubsection{Morphology} \label{subsubsect: morphologicalEvolution}
The basic morphology of the dense gas in our galaxy is presented in \autoref{fig:surfaceDensity_pBeta}, where we show the projected density of the three models. The lower limit of the colourbar is chosen such that the denser and colder regions are highlighted in the simulations. The rows represent different time snapshots as indicated in the panels, and the columns represent the runs with varying magnetisation. Starting with the first row ($t = 0.5\,T_{\rm rot}$), we can immediately see that the dense gas is dominantly present in the spiral arms. We can understand this simply by looking at the relevant drivers of the gas dynamics in the system. Since the rotation curve that we use is the analytical solution to the dark matter potential, it is the self-gravity, magnetic fields, the external spiral potential, and the equilibrium cooling/heating that drive the time evolution of the gaseous disc. The spiral potential funnels the gas towards its minima. This forms the dense spiral arms in the presence of self-gravity and cooling, while the magnetic fields oppose this by magnetic pressure ($P_{\rm mag} = B^{2}/ 8 \pi$).  Moreover, since the initial parameters of the galaxy are such that it is stable to toomre, thermal and swing instabilities, the spiral arms dominates the presence of dense gas in our galaxies. We can see the effect of magnetic fields, even as the spiral arms are forming, when we compare the three runs in the first row of \autoref{fig:surfaceDensity_pBeta}. The spiral arms are visibly more diffuse with increasing magnetisation due to the additional pressure support of the magnetic fields. As expected, this effect is more pronounced in the $\beta = 10$ case, since the $\beta$ of the gas is considerably larger in that case than in the $\beta = 50$ run, where the magnetic fields are sub-dominant in the initial phases of the evolution.  

We begin to see major differences between the evolutionary paths of the spiral arms between the three runs after they form, as seen in the next panel at $t = 0.75\,T_{\rm rot}$, where the arms start fragmenting into a beads-on-a-string pattern for the hydro and $\beta = 50$ runs, while on the other hand, the spiral arms are diffusing away in the $\beta = 10$ case. This then leads to stark differences at $t = 0.9\,T_{\rm rot}$, where the spiral arms that we see at $t = 0.75\,T_{\rm rot}$ have distinctly separated into clouds for the hydro and $\beta = 50$ cases. While they are diffused away in the $\beta = 10$ case. Moreover, we see secondary arm formation for all the three cases, visible on the inner face of the fragmented arms in the hydro and $\beta = 50$ cases, and in the absence of a fragmented arm, solitary in the $\beta = 10$ case. We continue running the $\beta = 10$ simulation for $t = 1.67\,T_{\rm rot}$, and observe that the disc goes through cycles of arm formation and diffusion, and that these arms never manage to fragment.



\begin{figure*}
    \centering
    \includegraphics[width=0.99\linewidth]{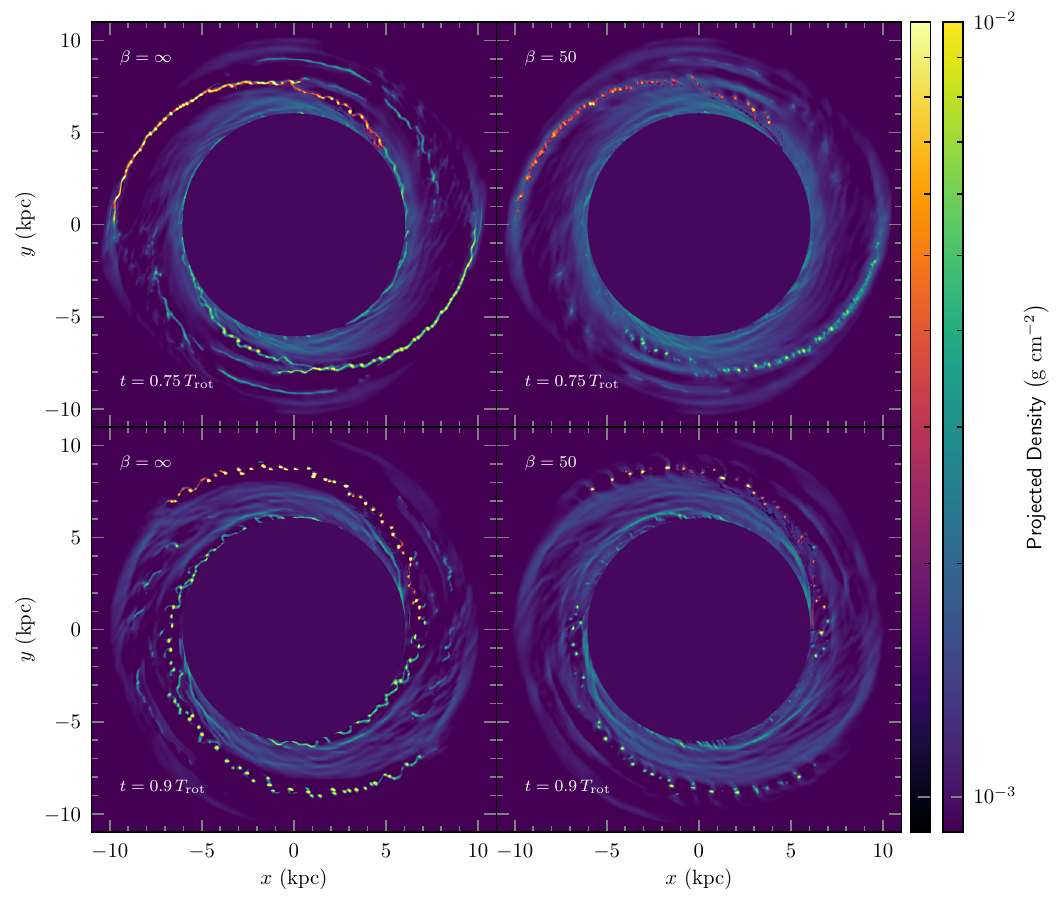}
    \caption{Projected density as in \autoref{fig:surfaceDensity_pBeta}, but only for the fragmenting cases ($\beta=\infty$ and $50$), with one of the arms traced and highlighted in a different colour scheme. We can see morphological differences in the manner in which the spiral arm fragments in the presence of magnetic fields.}
    \label{fig:surfaceDensity_0p75_spiralTraced}
\end{figure*}

Now we focus on the two runs where we see the spiral arms fragment, namely $\beta = \infty, 50$. In the \autoref{fig:surfaceDensity_pBeta} we can see morphological differences between the two runs at $t = 0.75\,T_{\rm rot}$, even though there are no notable differences at $t = 0.5\,T_{\rm rot}$, when the spiral arms form. The differences are better seen in the \autoref{fig:surfaceDensity_0p75_spiralTraced}, where we plot the projected density of the two runs and highlight the cells in the spiral arms with a different colour scheme to accentuate the difference. The two rows are at different times, analogous to the last two rows in \autoref{fig:surfaceDensity_pBeta}. We trace one of the arms using a friends of friends (FoF) algorithm with a linking length of $60\,\si{pc}$, using cells that are on the verge of being thermally unstable with $n_{\rm thresh} = 0.9 \, n_{\rm crit}$, where $n_{\rm crit} = 1\,\si{cm^{-3}}$ is the critical density of the thermally unstable medium. In the first panel, at $t = 0.75\,T_{\rm rot}$, we can see that the $\beta = 50$ run has the dense structures 
all radially elongated, while the $\beta = \infty$ one has roughly spherically-shaped over-densities. The second difference is that we see wiggles reminiscent of Kelvin-Helmholtz instabilities, connected with continuously with the gas present in the spiral arms in the hydro run. On the other hand, the spiral arms separate out into distinct clouds without any Kelvin-Helmholtz-like structures in the runs with magnetic fields. There are differences that persist even at late times. At $t = 0.9\,T_{\rm rot}$ in \autoref{fig:surfaceDensity_0p75_spiralTraced} we see a larger number of fragments more closely packed in the hydro run compared to the magnetic-field runs. This suggests a different mode of fragmentation in the presence of magnetic fields compared to without magnetic fields.

\subsubsection{Timescales} \label{results_subsect:timeEvol}

\begin{figure}
    \centering
    \includegraphics[width = 0.9\linewidth]{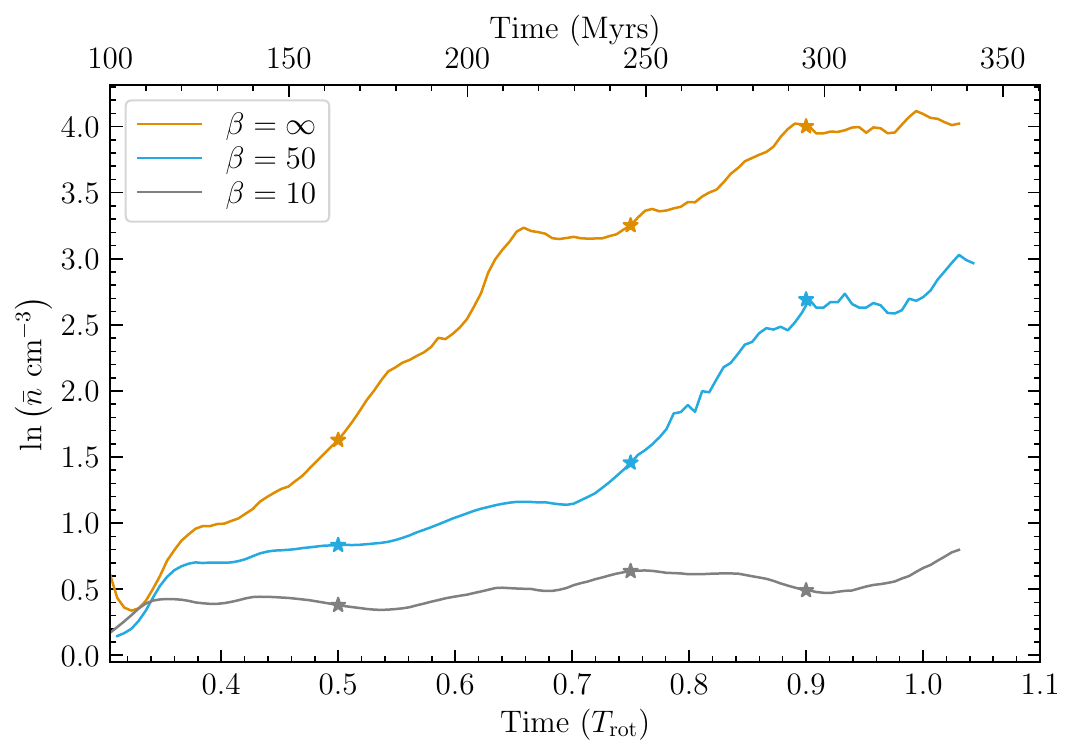}
    \caption{Time evolution of the mass-weighted mean density ($\bar{n}$) of all the cells with $n\geq 1\,\si{cm}^{-3}$. The bottom abscissa shows the time normalized by the rotation period, and the top abscissa shows the absolute time. The stars mark the times at which we show the disc morphology in the \autoref{fig:surfaceDensity_pBeta}. We can see that the $\beta = 10$ case does not have any appreciable mass above the threshold density of $1\,\si{cm^{-3}}$, as opposed to the other two cases. After $t = 0.75\,T_{\rm rot}$ we see the mean density of the $\beta = 10$ run decreases slightly as the spiral arms diffuse away, while it rises in the other two cases as the arms fragment into denser clouds.}
    \label{fig:time_evol_dense_gas}
\end{figure}

A more quantitative picture of the time evolution of the dense gas is presented in \autoref{fig:time_evol_dense_gas}. Here, we show the time evolution of the mass-weighted average density of the gas above the density threshold for the thermally unstable regime, $n>1\,\si{cm^{-3}}$. This is representative of gas that has the potential to become denser since it is in the thermally unstable regime. As expected, we find that this traced, dense gas, is predominantly present in the spiral arms. We visually confirm this by following the evolution of the disc with $10^5$ passive tracer particles that are initialised at $t=0$ in our region of interest (movies to be found in the supplementary material). The three solid lines in \autoref{fig:time_evol_dense_gas} are the three runs with different initial magnetisation. The stars indicate the times at which we show the disc in \autoref{fig:surfaceDensity_pBeta}. We can break down the time evolution of the dense gas into two phases - one where the spiral arms grow and the second one where it either fragments into clouds or diffuses away. The spiral arm growth phase starts at $t \simeq 0.3\,T_{\rm rot}$ $( 100\,\si{Myr})$ for all the runs and it lasts till $t \simeq 0.64\,T_{\rm rot}$ $(210\,\si{Myr})$ for the hydro and $t \sim 0.70\,T_{\rm rot}$ $(230\,\si{Myr})$ for the $\beta = 50$ run. In the $\beta = 10$ case on the other hand, even though the spiral arms have an appreciable amount of gas around $n\sim 1\,\si{cm^{-3}}$, as seen in the column density plots in \autoref{fig:surfaceDensity_pBeta}. The arms never manage to get appreciably dense. In the next phase of the  evolution, the $\beta = 10$ run repeatedly forms transient arms that quickly diffuse after their formation. This phase is seen as small crests and troughs in the \autoref{fig:time_evol_dense_gas} at $t \simeq 0.54, 0.74, 0.9 \, T_{\rm rot}$. In contrast, the spiral arms in the other two cases fragment. This is visible as a sudden change in the slope of $\ln \bar{n}$ in the same figure \autoref{fig:time_evol_dense_gas}, which marks the end of the spiral arm growth phase. 
During this phase, the average density rises at a faster rate in the $\beta = 50$ case when compared with the hydro run. For similar time intervals, from $t = 230$ to $t = 290$~Myr, the density rises by a factor of just $2.2$ for the hydro run, in contrast to a factor of $5$ in the magnetic run. This is seen in the steeper slope of the former compared to the latter in \autoref{fig:time_evol_dense_gas}. Looking at the tracer particle movies, we see that the hydro run separates out into distinct clouds at around $t \simeq 290 \,\si{Myr}$, while the $\beta = 50$ run does it much quicker by $t \simeq 270 \,\si{Myr}$.

\subsection{Cloud Separation and Fragmentation Modes}\label{result_subsec:cloud_spacings_and_fragmentation_modes}

Here we discuss how the separation (spacing) of the clouds that form in the spiral arms of the $\beta = 50$ case and the hydro case differ from each other. This gives us insight into the effects of the magnetic fields and on the fastest growing mode of the spiral instability. 

\subsubsection{Extracting the cloud spacings} \label{result_subsect: extracingSpacings}
We quantify the separation of the clouds and the wavelength of the unstable modes by using 1D Fourier analysis. We do this on the projected density binned along the length of the spiral arms. We first construct a spiral arm mask using analytical functions and then use this spiral coordinate to bin the density in preparation for the Fourier transformation. 

\begin{figure}
    \centering
    \includegraphics[width = 0.85\linewidth]{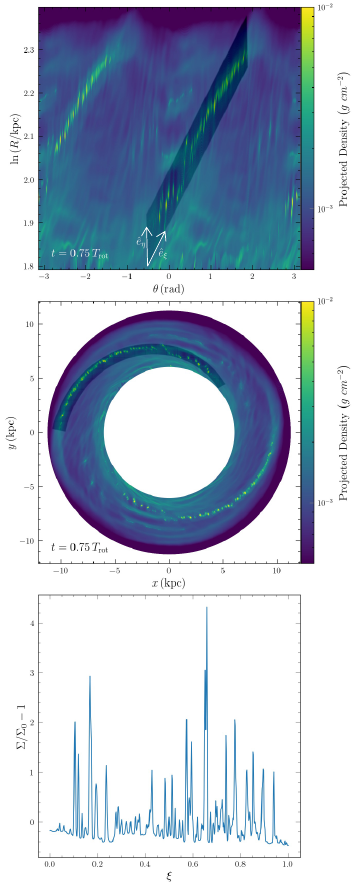}
    \caption{The first two panels show the spiral arm mask we use in the $\left ( \ln{R}, \theta \right )$ plane and in the $\left ( x,y \right )$ plane, respectively. The first panel also shows the unit vectors $\left ( \hat{e}_{\xi}, \hat{e}_{\eta} \right )$. The third panel shows the binned relative projected density $\Sigma/\Sigma_{0} - 1$ along the arm. We can see the clumps as peaks in relative density in the third panel.}
    \label{fig:binned_density_spiral_coordinate}
\end{figure}

Similar to \cite{roberst_69} we define  
\begin{align} \label{eqn: spiral_coordinate_xi}
    \xi &=  \ln{ \left (R/R_{0} \right )} \sin{p}  +  \theta \cos{p}, \\
    \eta &= \ln{ \left (R/R_{0} \right )} \cos{p}  -  \theta \sin{p},\label{eqn:spiral_coordinate_eta}
\end{align}
as the spiral arm coordinates, where $ R_{0} = 1\,\si{kpc}$ is the scaling and $p$ is the pitch angle of the spiral arm under consideration. The $\left ( \xi, \eta \right )$  coordinates can be thought of as the $\left ( \ln{R}, \theta \right )$-plane rotated counter-clockwise by the angle $p$. Here $\xi$ is the coordinate along the spiral arm and $\eta$ is locally perpendicular to it. To define a spiral arm region of a certain thickness, we use a rectangular mask in the $(\xi, \eta)$-plane, where the thickness will be decided by the extent in the $\eta$ coordinate. Our spiral arm mask is shown in \autoref{fig:binned_density_spiral_coordinate}, where the first panel shows the projected density of the galaxy at $t = 0.75\,T_{\rm rot}$ in the  $\left ( \ln{R}, \theta \right )$ plane along with the masked region shaded and the unit coordinate vectors $\left(\hat{e}_{\xi},\hat{e}_{\eta}\right)$ on the bottom edge of the mask. The same plot is shown in the $\left ( x, y \right )$ plane in the second panel, and the third panel shows the binned projected density along the spiral arm (coordinate $\xi$). We construct the mask starting with an initial guess for the pitch angle, $p_{i}$, that determines the $(\xi, \eta)$ plane. Next, we draw the rectangular mask in this plane that is then a rectangle bounded via fixed values of lower edge $\left ( \ln{R_{0}}, \theta_{0} \right )$ and the maximum radial coordinate $\left (\ln{R_{\rm max}} \right )$ that we determine via visual inspection at the beginning. Now, we test the correctness of this initial rectangular mask by fitting a straight line to all the  $\left ( \rm ln R, \theta \right )$ coordinates of cells that lie in this mask weighted by their densities. If it were a good mask that covers all the dense regions then the slope of this line $m_{\rm line}$ will be close to $m_{i} = \tan{p_{i}}$. However, if it is not the case, we use it for the next iteration, where $p_{i+1} = \tan^{-1}(m_{i})$. For convergence we use an absolute tolerance of $ p_{i+1} - p_{i} = 0.05^{\circ}$. Once we have the mask, we simply bin the projected density on the coordinate along the spiral arm.



Our algorithm is similar to what \citet{gusev_efremov_2013} used on observational data, but with one key difference - here we explicitly allow for a certain thickness of the spiral arm mask in the direction perpendicular to its length, while \citet{gusev_efremov_2013} do not. Instead, they just fit the curve $\eta = const$ by selecting all the pixels that lie along the spiral arm by eye and determine the pitch angle $p$ of the spiral arm via a least square fit.  

The projected density on the coordinate along the spiral arm is seen in the bottom panel of \autoref{fig:binned_density_spiral_coordinate}. Here, the peaks in the column density are the encountered over-densities along the length of the spiral arm. We chose the number of bins such that there are $\geq 4$ cells in each bin. With this binned density, we finally take the 1D discrete Fourier transform, with the following convention:
\begin{equation}
    \hat{\Sigma}_{\rm sp}[k] = \frac{1}{N} \sum ^{N-1} _{m = 0} \Sigma_{\rm sp, rel}[m] \exp{\left (\frac{-2 \pi i k m}{N} \right )}, \quad k = 0, 1,...,N-1, 
\end{equation}
where $N$ is the total number of bins, $\Sigma_{\rm sp, rel}[m] = \Sigma_{\rm sp}[m]/\Bar{\Sigma}_{\rm sp} - 1$ is the relative-surface density along the spiral in the $m \rm th$ bin and $k$ is the wave number in units of $1/L_{\rm spiral}$, with $L_{\rm spiral}$ being the length of the spiral arm. We calculate $L_{\rm spiral}$ by averaging over the lengths of the inner and the outer edge of the spiral arms. The power is then taken to be, 
\begin{equation}
    P[k] = \hat{\Sigma}_{\rm sp} [k] \hat{\Sigma}_{\rm sp}^{*} [k],
\end{equation}

where $\hat{\Sigma}_{\rm sp}^{*} [k]$ is the complex conjugate of $\hat{\Sigma}_{\rm sp} [k]$. As a final step, we bin the resultant power spectrum in bins of length $k = 2$.


\subsubsection{Cloud separation}\label{result_subsect_cloudSpacings}
\begin{figure*}
\centerline{
\def\arraystretch{1.0}
\setlength{\tabcolsep}{1.3pt}
\begin{tabular}{cc}
\includegraphics[height=0.3\linewidth]{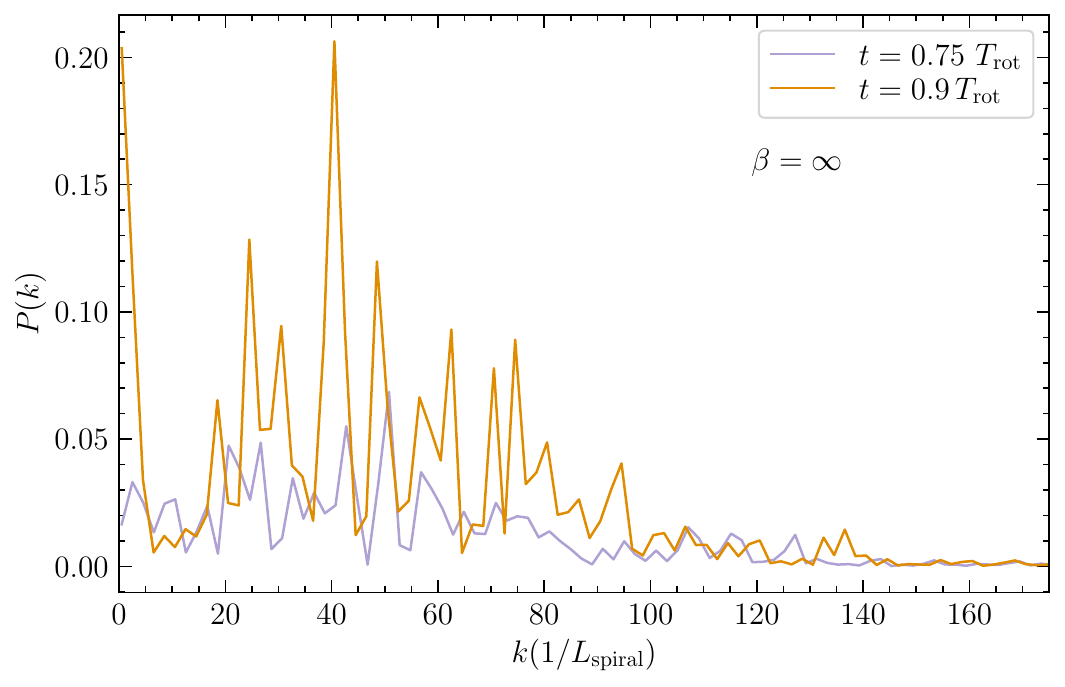}\label{subfig:power_spectrum_hydro} 
\includegraphics[height=0.3\linewidth]{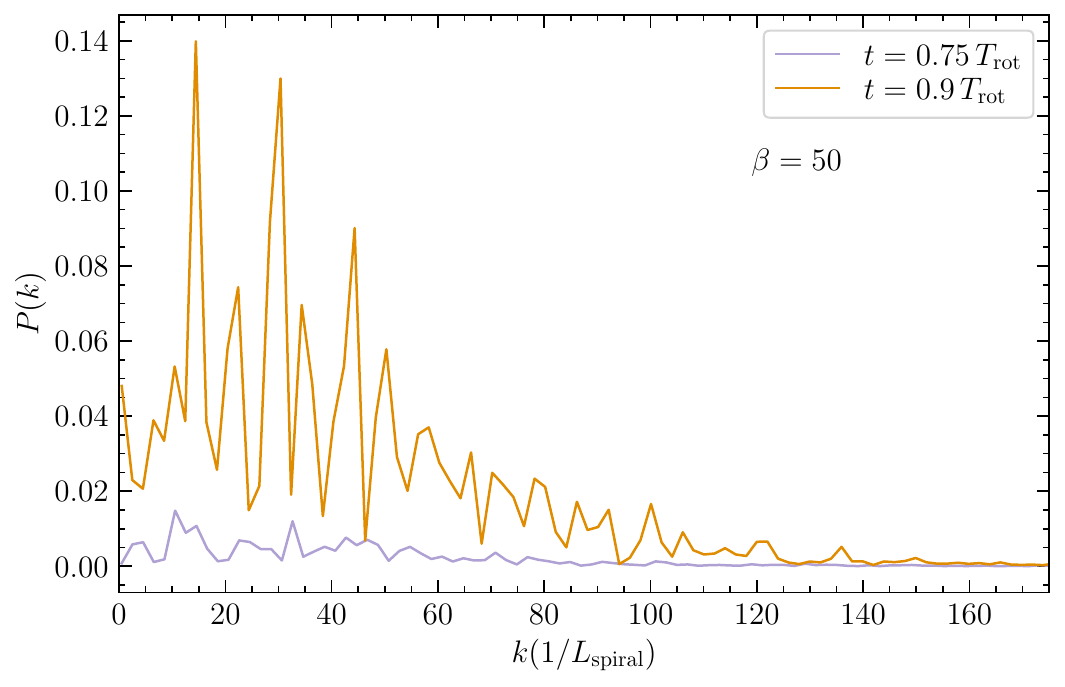}\label{subfig:power_spectrum_beta50}
\end{tabular}
}
\caption{Fourier transform of the projected density in the spiral arms for the $\beta = \infty$ and $\beta = 50$ runs. The two solid lines show the Fourier transforms at $t = 0.75, 0.90~ T_{\rm rot}$, analogous to the times shown in \autoref{fig:surfaceDensity_pBeta}. The wave number $k$ here is given in units of $1/L_{\rm spiral}$, i.e., $k=1$ corresponds to the full length of the spiral arm under consideration. We can see that the $\beta=50$ run has more power at lower $k$ (larger scales) than the hydro run, as we visually saw in \autoref{fig:surfaceDensity_0p75_spiralTraced}.}
\label{fig:power_spectrum}
\end{figure*}

We use the $1$D power spectrum of the projected density derived in the last section to quantify the separation of the clouds in the spiral arms. We calculate the power spectrum at $t \in \left \{0.75\,T_{\rm rot}, 0.90 \,T_{\rm rot} \right \}$ of one of the spiral arms as an example. These times correspond to the times shown in the bottom two panels of \autoref{fig:surfaceDensity_0p75_spiralTraced} and roughly indicate when the spiral arms are undergoing fragmentation and when the spiral arms have separated out into distinct clouds. The power spectrum for one of the arms is shown in \autoref{fig:power_spectrum}, where the left panel is for the hydro run and the right panel is for the $\beta = 50$ case. The power on the vertical axis is plotted against the wave number $k$, which has units of $1/L_{\rm spiral}$, where $L_{\rm spiral}$ is the length of the spiral arm . 
We can see the rise in the power from $t = 0.75 \, T_{\rm rot}$ to $t = 0.90 \,T_{\rm rot}$ as the arm fragments and the clouds separate out and become denser. This naturally gives us a higher signal, as the $\Sigma_{\rm sp, rel}$ grows and the spacings become more distinct. The hydro case has more power at early times, compared to the $\beta = 50$ run, as expected, since it starts fragmenting $t\sim 20\, \si{Myr}$ earlier than the latter (c.f. Fig.~\ref{fig:time_evol_dense_gas}). At later times, $t = 0.9\,T_{\rm rot}$, we see distinct multiple peaks with similar power, where majority of the power resides on larger scales with slight differences, for both the runs. For the hydro run this is for $k\lesssim 80$ ($l \gtrsim 260 \,\si{pc}$), while for the magnetic case we have $k \lesssim 50$ ($l\gtrsim 390\,\si{pc}$). 

The power spectrum also gives us insights into the regularity of the separation of the clouds in this particular arm. At $t = 0.90\,T_{\rm rot}$ the hydro run has a major peak at $k = 40$ ($l \sim 500\,\si{pc}$). Similarly, in the $\beta = 50$ case, we see a major peak on larger scales at $k = 30$ ($l \sim 650\,\si{pc}$). Both of these correspond to the total number of clouds that we see in the respective arms. This signal in the Fourier analysis indicates that the adjacent separations of the clouds are regular. The magnetic case also has a major peak present at $k = 14$ ($l \sim 1.4 ~\si{kpc}$ ), which corresponds to the number of brighter clouds along the spiral arm, also visible as the taller spikes in the bottom panel of \autoref{fig:binned_density_spiral_coordinate}.

An interesting feature in the Fourier transform common to both the runs, is the existence of major peaks that come in pairs of multiples of $2$. For example, we have $k \sim (24, 48), (30, 60), (40, 80)$ in the hydro case and $k \sim (14, 30), (22, 44), (34, 70)$ for the $\beta = 50$ case. This feature could be a combination of two things - the first is the physical presence of such modes due to the nature of the spiral arm instability, and/or the second is due to the asymmetry in the negative and positive values present in the function, $\Sigma_{\rm sp, rel}[m] = \Sigma_{\rm sp}[m]/\Bar{\Sigma}_{\rm sp} - 1$, of which we take the Fourier transform. In the latter case we expect a single peak of the dominant mode followed by even harmonics with a decreasing power-law amplitude of the subsequent peaks. Looking at the lower panel of \autoref{fig:binned_density_spiral_coordinate}, we can see that the function, $\Sigma/\Sigma_{0}$ - 1, is indeed asymmetrical, showing that the voids between the clouds are not as under-dense as the clouds are over-dense. However, apart from the dominant modes at low $k$, the higher $k$ modes all have comparable amplitudes, and thus we consider them to be physical modes present in the system. 

\begin{figure*}
    \centering
    \includegraphics[width = 0.95\linewidth]{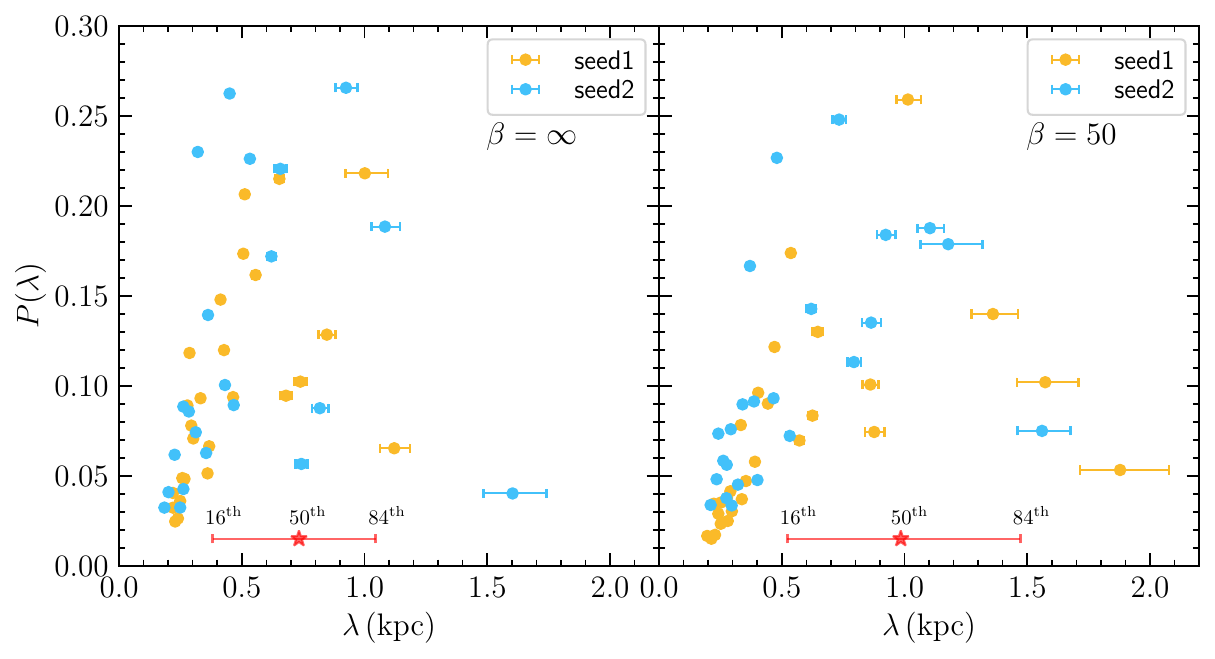}
    \caption{Power ($P(\lambda)$) of the extracted peaks of the Fourier transforms of the projected density along the masked spiral arms at $t = 0.9\,T_{\rm rot}$, plotted against the physical scale in $\rm kpc$. The two panels show the hydro and the $\beta = 50$ cases, respectively. The two runs with different random seeds are shown in yellow and blue, respectively. In each case, we have included the peaks found for both the spiral arms. The star marks the 50th percentile and the error bars indicate the range from 16th to 84th percentile of the peaks binned in $0.2\, \si{kpc}$ bins. It increases from $0.73^{+0.31}_{-0.36}$~kpc in the hydro case to $0.98^{+0.49}_{-0.46}$~kpc in the magnetic case. However, peaks with most power in both the runs are present on length scales of $\sim 0.5 - 1.0$ \si{kpc}.}
    \label{fig:spacings_physicalSpace}
\end{figure*}

We test whether the differences observed in one of the spiral arms between the hydro and the magnetic cases can be generalised. For this, we run an identical pair of simulations for both the hydro and $\beta = 50$ cases with a different random seed for the initial turbulence. We then also use both the spiral arms of each simulation in our analysis. As done before, we calculate the power spectrum of the projected density along each of the arms. We show the peaks of the power spectrum in \autoref{fig:spacings_physicalSpace}. Here, the error bars indicate the FWHM with a minimum value equal to the binning length $k = 2$, with respect to the physical scale in \si{kpc} on the horizontal axis. The left panel is for the hydro run and the right panel for the $\beta = 50$ run. The different colours are for the runs with different initial turbulent seeds. The star and its error bars indicate the 50th (median) and the 16th to 84th percentile range after the points are binned in bins of $0.2$~kpc. 

We see from the \autoref{fig:spacings_physicalSpace} that the trends we observed in one of the spiral arms carry over to the general case as well. The power in both the cases exists on large scales, with the hydro run having major power over $l\gtrsim300\,\si{pc}$ and the magnetic case having it on $l\gtrsim400\,\si{pc}$. We also see many significant peaks that are a multiple of $2$ of a lower $k$ mode in both the magnetised as well as the hydro case. The rise in the cloud separation and unstable modes seen in the Fourier transform of one arm is also seen as a general trend of the presence of magnetic fields. Moreover, we also see that the power rises more steeply on small scales and also falls sharply after $1\,\si{kpc}$ in the hydro case. For the magnetic-field case the rise in the power is shallower on smaller scales and spreads towards larger scales till $1.5~\si{kpc}$. This is reflected in the percentiles of the distribution of the peaks, where both the median and the $16$th to $64$th percentile range of the hydro case rises from $0.73^{+0.31}_{-0.36}$~kpc to $0.98^{+0.49}_{-0.46}$~kpc in the magnetised case. 


\subsection{Effects of magnetic fields on the physical properties of spiral arms}\label{result_subsect_spiral_arm_properties}

\begin{figure*}
    \centering
    \includegraphics[width=0.99\linewidth]{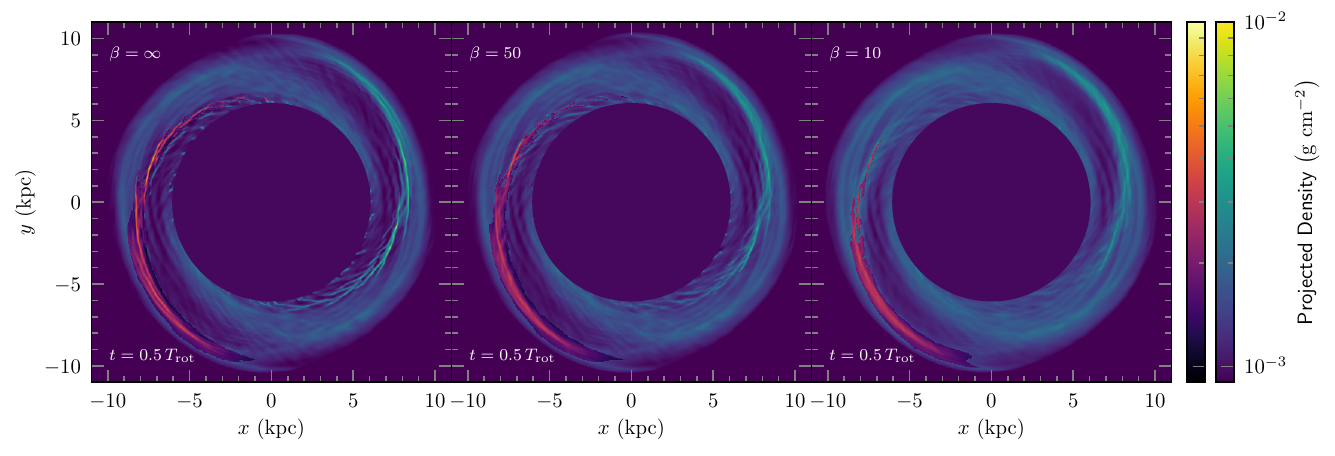}
    \caption{Projected gas density of our models at $t = 0.5 \,T_{\rm rot}$ along with the traced spiral arm highlighted with a different colour scheme. We can see that while the spiral arms in the hydro and $\beta = 50$ cases are morphologically similar to one another, there are noticeable differences with the $\beta = 10$ case.}
    \label{fig:surfaceDensity_0p50_spiralTraced}
\end{figure*}

As we have seen, the presence of magnetic fields, regardless of their strength, influences the evolution of the spiral arms. The magnetic fields themselves, are also expected to evolve with the gas in our simulations. In order to gain insight into this, we explore the physical properties of the spiral arms in the three different runs with different initial magnetisation before they fragment or diffuse away. We use one of the spiral arms in each simulation, since we find that they have very similar physical properties. As done in \autoref{fig:surfaceDensity_0p75_spiralTraced}, we trace the gas in the spiral arm under consideration at $t = 0.5 \,T_{\rm rot}$ using a friends of friends algorithm with a linking length of $l = 60 \, \si{pc}$, which is the approximate cell length of the surrounding warm neutral medium around the spiral arms. In addition, we also use a density threshold of $n = 0.9\,\si{cm^{-3}} $, which is similar to the critical density of the thermally unstable medium. The traced spiral arms is presented in \autoref{fig:surfaceDensity_0p50_spiralTraced}, where we show the projected density of the three runs along with the traced spiral arms highlighted in a different colour scheme. We use these traced regions for all the properties we report in this section. 

\begin{figure*}
\centerline{
\def\arraystretch{1.0}
\setlength{\tabcolsep}{1.3pt}
\begin{tabular}{ccc}
\includegraphics[height=0.22\linewidth]
{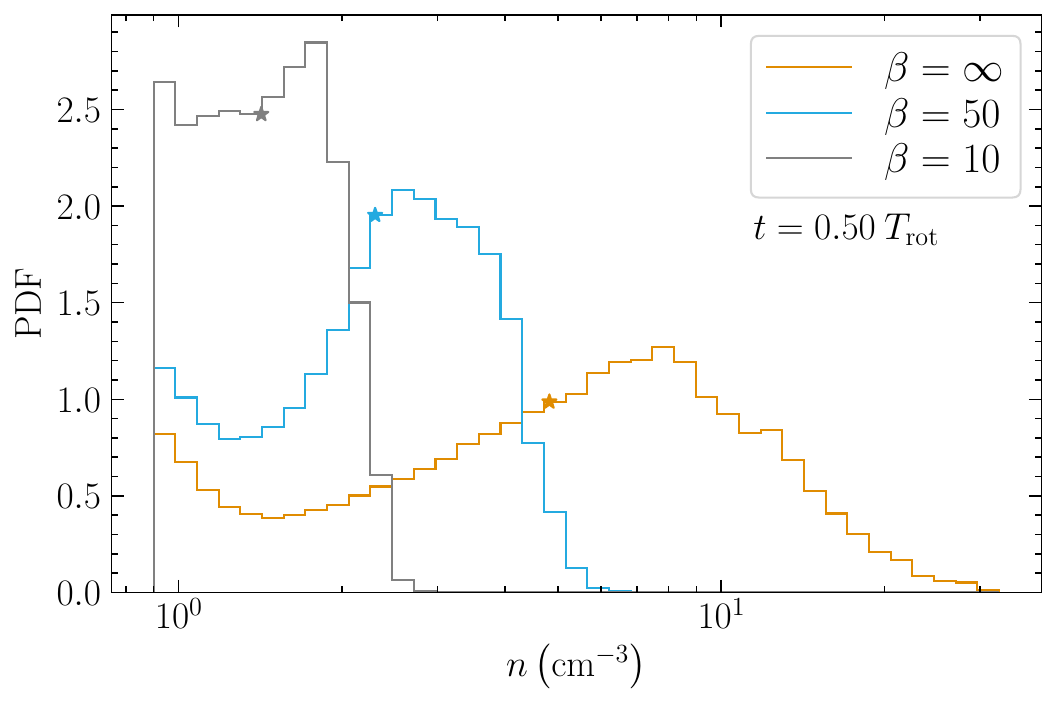}\label{subfig:density_pdf} 
\includegraphics[height=0.22\linewidth]{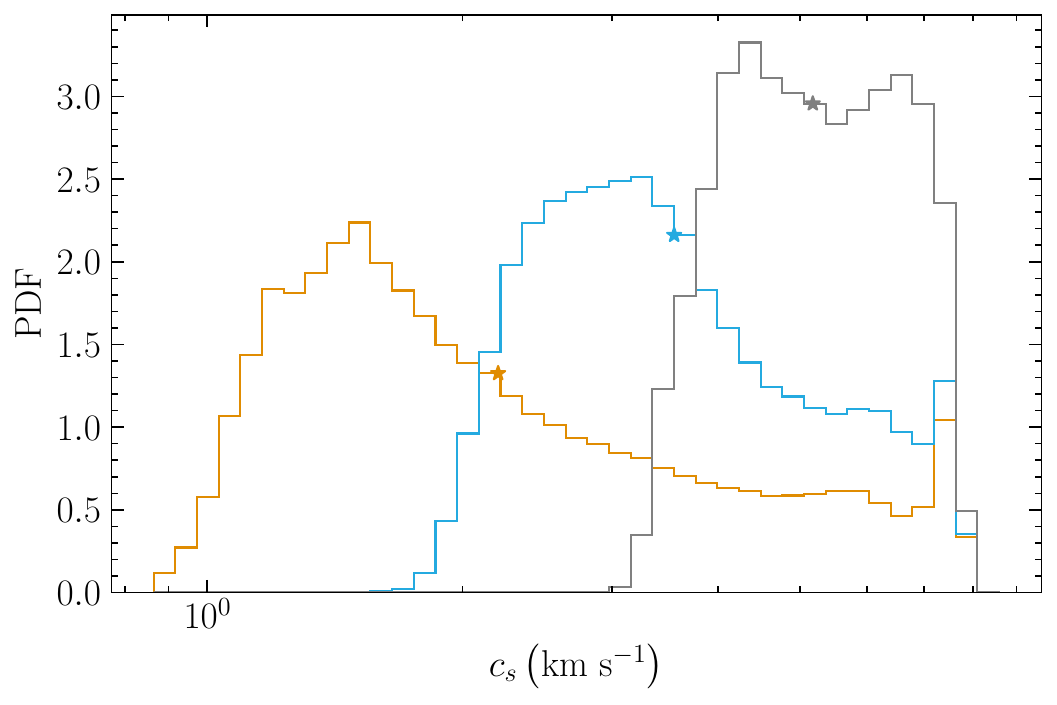}\label{subfig:soundSpeed_pdf}
\includegraphics[height=0.22\linewidth]{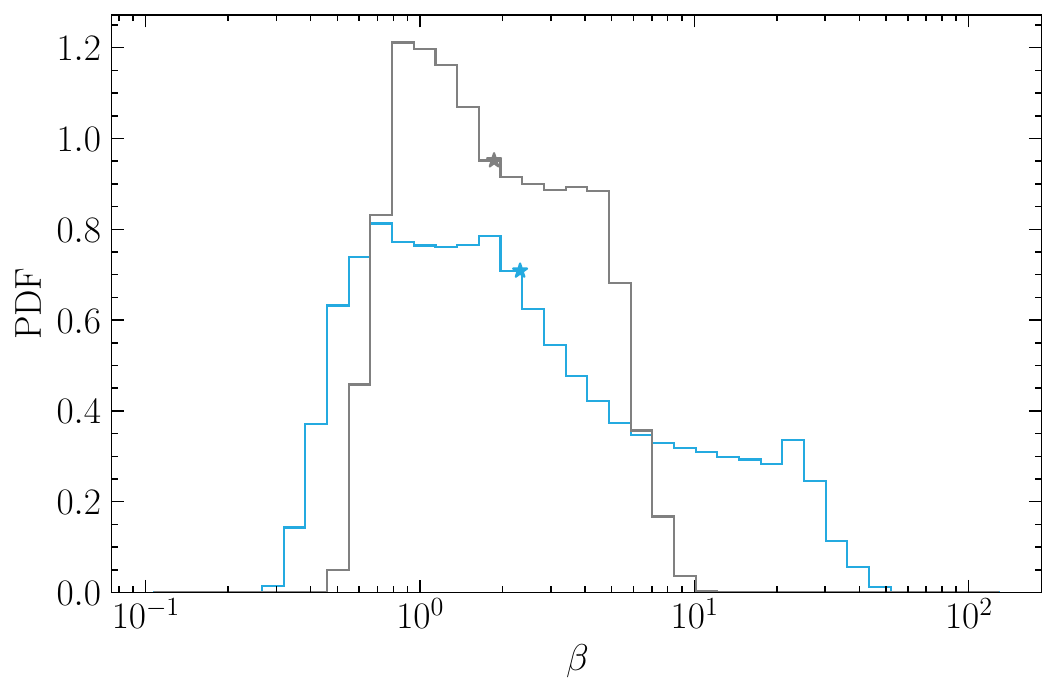}\label{subfig:pBeta_pdf}
\end{tabular}
}
\caption{Mass-weighted probability density functions (PDFs) of the $\log_{10}$ of gas density (left panel), sound speed (middle panel), and cell-by-cell plasma-beta (right panel) of the traced spiral arms at $t = 0.5 \, T_{\rm rot}$. The stars mark the mass-weighted average of the $\log_{10}$ of the quantity in each case. In the left panel, we see that the magnetic fields make the spiral arms more diffuse. The middle panel shows how the gas gets colder with increasing density as the cooling sets in. In the right panel we see that the plasma-beta of the magnetic runs have decreased significantly in the arms (compared to the simulation starting value of $\beta$), especially for the $\beta = 50$ case, where the plasma-beta has risen from $50$ to an average of $2.3$, becoming comparable to the $\beta = 10$ case. 
}
\label{fig:pdf_0p50}
\end{figure*}

The total mass present in the arms is similar, i.e., $\log_{10} \rm (M/(M_\odot)=7.90$, $7.85$ and $7.71$ for the hydro, $\beta = 50$, and $\beta = 10$ case, respectively. However, other physical properties vary systematically between them. We can see this in the \autoref{fig:pdf_0p50} where we show the mass-weighted probability density functions (PDFs) of the $\log_{10}$ of density in the left panel, the sound speed in the middle panel, and the cell-by-cell plasma-beta of the runs in the right panel. The stars on the histograms mark the average mass-weighted quantities in the respective vertical axis. In the left-most panel, as we saw in \autoref{results_subsect:timeEvol}, we see that the spiral arms get more diffuse with increasing magnetisation. As seen in the mean density, that decreases by a factor of $\sim 2$, possibly due to the additional opposing magnetic pressure. We find that the majority of the gas in the $\beta = 10$ case is present in the density range $\simeq 1$--$3\, \si{cm^{-3}}$ which is thermally unstable. Despite this, it never manages to get denser due to the opposing magnetic pressure. We can also see this effect in the middle panel, where the gas in the hydro and the $\beta = 50$ cases have a lower sound speed when compared to the $\beta = 10$ case, showcasing that the gas has already cooled in the former two cases, while the gas remains warm and close to its initial temperature in the latter. This is also reflected in the scale heights of the spiral arm, where it is found to be $ 50.2 \pm 2.2\, \si{pc}$ in the $\beta = 50$ case and $98.6 \pm 3.9\, \si{pc}$ in the $\beta = 10$ case. The details of the scale height calculation can be found in \aref{Appendix:scale_height_estimation}. Thus the presence of magnetic fields largely makes the spiral arms more diffuse and hotter. 

Interestingly, even though we start with a factor of $5$ difference in the initial plasma-beta of the two magnetised runs, we note that they have similar values in the spiral arms. We see this in the right-most panel of \autoref{fig:pdf_0p50}, where the plasma-beta of the run with initial $\beta = 50$ has reduced by a factor of $\sim 25$ to an average value of $\sim 2$, and even has a non-negligible fraction of cells with values $\leq 1$, where the magnetic fields dominate over the gas pressure. The $\beta$ in $\beta = 10$ run reduces by a factor of $5$
to reach similar values as in the $\beta=50$ case. This shows that the magnetic fields, even though not dynamically important at the beginning of the simulation do become important later on in the spiral arms. This increase in field strength could be due to the combined effects of field tangling, adiabatic compression, and cooling, or the presence of a dynamo \citep{Federrath2016jpp}. These effects are much more pronounced in the $\beta = 50$ case, where the gas compresses and cools in the absence of dynamically significant magnetic fields, when compared to the $\beta = 10$ case, where they oppose compression and cooling. Thus, in the $\beta = 50$ case, we get magnetic fields that are dynamically significant "after" the spiral arms have become dense enough. This results in fragmentation in the presence of magnetic fields, and changes the nature of the instability when compared with the hydro case. We discuss this in detail in the next section.

\section{Discussion} \label{sec:Discussion}
Here, we discuss the physical effects of magnetic fields on the spiral arm instability and compare them with previous theoretical and observational studies. First, we focus on the stabilising effect in \autoref{subsect:stabalization}, their destabilising effects in \autoref{subsect:destabalization}, and then go on to discuss the cloud separation and unstable modes in \autoref{subsect:lengthScales}. Lastly, we point out some caveats of our work in \autoref{subsect:caveats}.

\subsection{Stabilisation} \label{subsect:stabalization}

We find that moderate initial magnetic fields with initial $\beta = 10$, can stabilise the spiral arms against fragmentation. As seen in \autoref{fig:pdf_0p50}, the spiral arms that form in this case are more diffuse and hotter compared to the other cases (hydro and $\beta=50$ models) where they fragment. This is mainly due to the increased magnetic pressure of the B-fields that opposes the gas from getting any denser. This inhibition of compression due to the additional magnetic pressure agrees well with other global disc galaxy simulations \citep{dobbs_price_2008, khoperskov_2017, bastian_2019}. However, we note that our value of $\beta = 10$ for stabilisation is higher than the ones observed in other studies --  $\beta \leq 0.1$ \citep{dobbs_price_2008} and $\beta \leq 1$ \citep{khoperskov_2017}. As pointed out before, our models are gravitationally stable initially, and also have low shear and a warm medium.

For weak magnetic fields, $\beta = 50$, even though we see that the spiral arms fragment, they do so in a different morphological manner than in the hydro run. The KHI-like wiggles as seen in the hydro case \citep[reported in][]{wada_koda_2004, wada_2008, sormani_sobacchi_2017, mandowara_sormani_2021}, are not present in the weakly magnetised simulation. This is because the magnetic field in the spiral arms becomes dynamically important, with $\beta_{\rm arm} \sim 2$, which then opposes the wiggles due to magnetic tension. This stabilising effect of magnetic fields, indeed, has been reported for magnetic fields of near equipartition strengths in both global 2D \citep{shetty_ostriker_2006} and local 3D simulations \citep{kim_ostriker_2006}. 

\subsection{Destabilisation} \label{subsect:destabalization}

For the case with weak initial magnetisation, the magnetic fields rise to equipartition levels within the arm (c.f., right-hand panel of Fig.~\ref{fig:pdf_0p50}). However, instead of stabilising the arm, as we expect from the additional magnetic pressure, the arms still fragment. They do so by clearing out the gas within the arm into clouds $\sim 20 ~\si{Myr}$ before the hydro case (see \autoref{results_subsect:timeEvol}). We attribute this to the possible presence of the Parker instability within the spiral arms. Parker instability \citep{parker_1966, parker_1967_I, parker_1967_II, parker_1968_I, parker_1968_II} arises in a magnetised plasma present in a stratified medium akin to a disc galaxy. A fluid element with a small magnetic-field over-density in the disc becomes more diffuse due to the added magnetic pressure, and will tend to rise upwards. Since the medium is stratified, after rising, the fluid element looses more gas to adjust to the decrease in the ambient pressure, thus becoming lighter and more unstable \citep{shukurov_subramanian_2021}. This eventually results in a characteristic magnetic field structure of regular hills and valleys above and below the galactic plane, where clouds are expected to form in the valleys of the magnetic-field lines. In our simulations, we suspect that the initial magnetic over-densities are naturally provided by the spiral arms. 

To test the plausibility of the presence of Parker instability, we estimate the expected growth rates and the length scales from linear theory for the gas in the spiral arms before they fragment, at $t = 0.50\,T_{\rm rot}$. Taking the average physical properties of the spiral arm presented in \autoref{result_subsect_spiral_arm_properties}, with $c_{s, \rm arm} = 3.55\,\si{cm}\,\si{s^{-1}}$, $\beta_{\rm arm} = 2.32$, $H_{\rm arm} = 50.2\,\si{pc}$, and $\gamma_{e} = 1$, we find an inverse growth rate of $\tau \simeq 33\,\si{Myr}$, and the wavelength of the fastest growing mode $\simeq 600\,\si{pc}$ \citep[see eqs.~17 and 22 in][]{mouschovias1996}. As a result of cooling present in our simulations, the gas is also expected to cool in the valleys as it gets denser \citep{hanasz_kosinki_2006, mouschovias_2009}. This, along with self-gravity is expected to increase the growth rates of the instability, which makes both the length scales and time scales remarkably close to the ones we observe in the spiral arms. From \autoref{fig:time_evol_dense_gas}, we can roughly estimate $\tau_{\rm arm} \sim 30\,\si{Myr}$, and as we saw in \autoref{result_subsec:cloud_spacings_and_fragmentation_modes}, the cloud separation in the $\beta = 50$ runs are $\simeq 650\,\si{pc}$.

\begin{figure*}
    \centering
    \includegraphics[width = 0.90\textwidth]{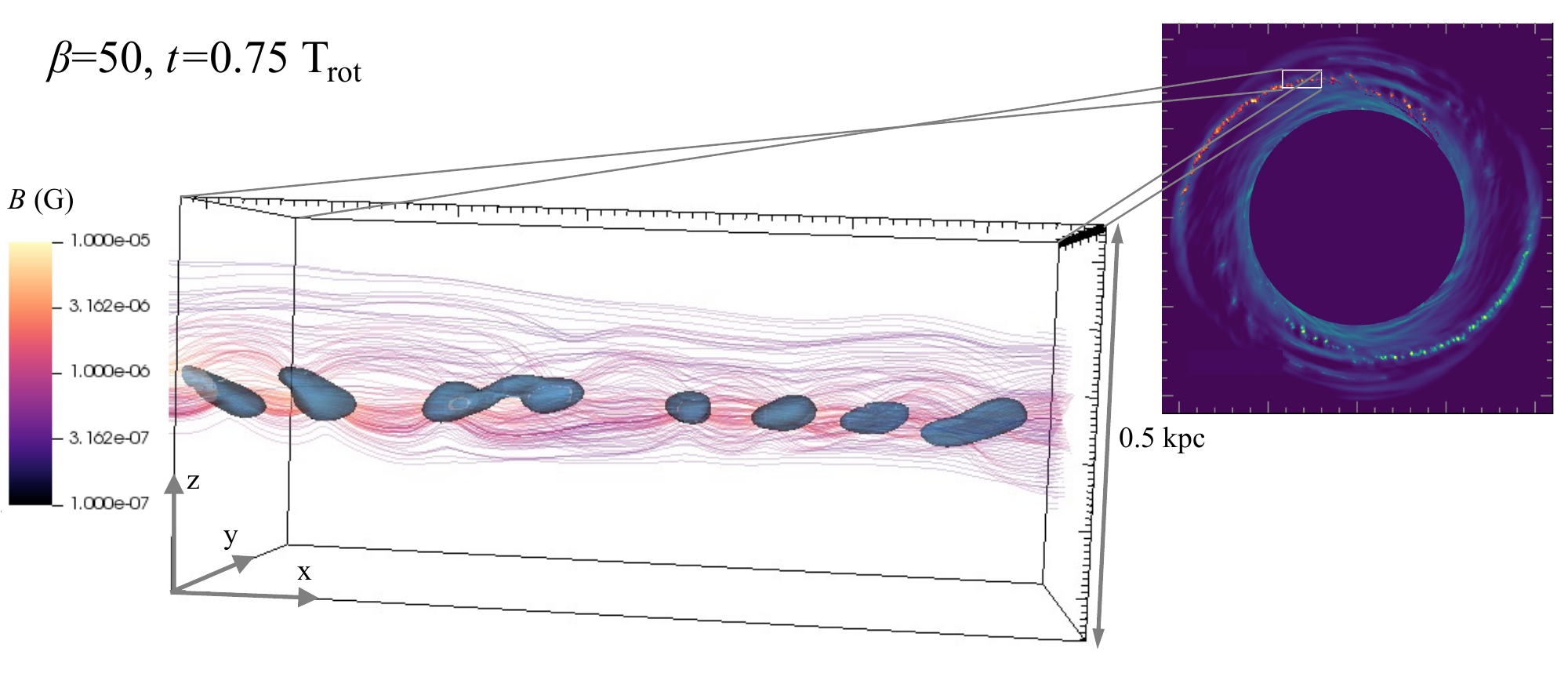}
    \caption{Magnetic field lines in a section of one of the spiral arms for the $\beta = 50$ simulation at $t = 0.75 \,T_{\rm rot}$. The magnetic field lines, coloured by their magnitude, were initialised on the right face (yz) of the region, in a circular plane of radius $250$~pc, with its z-axis aligned with that of the clouds. The clouds are shown in blue iso-contours with a threshold density of $\sim 2$ times the critical density of the thermally-unstable medium. We can see evidence for Parker loops that rise above and below the plane of the clouds to scales of $\sim 100$~pc, where the clouds lie in the valleys made by the field lines.}
    \label{fig:parker_instability_3D}
\end{figure*}

In addition to this, we see the characteristic magnetic field morphology associated with the Parker instability in our spiral arms. A section of this characteristic field structure is shown at $t =  0.75\, T_{\rm rot}$ in \autoref{fig:parker_instability_3D}. To produce this graph, we initialise the magnetic field lines at one $yz$ face of the three-dimensional box in a circular plane of radius $250$ pc, coloured by the field strength. The clouds are solid iso-contours with a density of $2\,\rm cm^{-3}$, which is about twice the critical density of the thermally-unstable medium. We can see the magnetic-field lines rising above and below the plane of the spiral arm on scales of $\sim 100\,\si{pc}$. Since our thermally-unstable medium is in the range $1-10\,\si{cm^{-3}}$, the gas predominantly cools in the magnetic valley as expected. Our results are in agreement with \cite{mouschovias_2009}, who found that the Parker instability in unison with the thermal instability can lead to the formation of dense clouds in sections of spiral arms. In contrast to the $\beta=50$ case, the seeds of the instability are never provided for the $\beta = 10$ case, since the spiral arms keep dispersing away before they get dense enough to fragment when the field is too strong.

\subsubsection{Comparison to observations}
One nearby galaxy, NGC $628$, recently has been found to have evidence of magnetic Parker loops along one of its spiral arms in RM synthesis maps \citep{mulcahy_beck_2017}. These loops are roughly coincident with the regularly spaced star-forming regions that were studied in \cite{gusev_efremov_2013}. A similar pattern was seen in the NW arm of M31, where the wavelength of the Parker loops, $\sim 2.3~\si{kpc}$ \citep{beck_1989}, was found to be twice the separation of the regularly spaced star-forming regions found along the arm \citep{efremov_2009, efremov_2010}. These are encouraging signs of the presence of the Parker instability. However, to draw firm conclusions, more detailed analyses of a wider sample of galaxies that exhibit this regular spacing of star-forming regions along their spiral arms is needed. 
\subsubsection{Comparison to simulations}
Linear stability analysis and local 2D simulations have observed a reduction in the growth rate of the spiral instability by a factor of $4$, in the presence of equipartition magnetic fields in comparison to the hydrodynamical case \citep{kim_wong_kim_2014}. This is in contrast to our results, where we see an increase in the growth rates. This discrepancy could be due to their 2D approximations, where they do not capture the onset of the Parker instability as observed in our simulations. \cite{kim_ostriker_2006} observed magnetic destabilisation in their local 3D simulations of the spiral shock fronts, but attributed it to the magneto-Jeans instability (MJI), as they did not observe the characteristic Parker loops. We think this is not the case in our simulations, since the spiral arms separate out into clouds when the arms are at considerably low densities ($n_{\rm av} \sim 5~\si{cm^{-3}}$). It was also later argued that \cite{kim_ostriker_2006} had insufficient box size perpendicular to the plane of the disc to find the Parker modes \citep{mouschovias_2009}.

\subsection{Cloud separation} \label{subsect:lengthScales}

\subsubsection{Comparison to observations}
Out of many others, only four spiral galaxies, namely NGC $628$ (M$74$), NGC $895$, NGC $5474$, NGC $6946$, have been analysed in detail for the separation of the regularly spaced star-forming regions, using a method similar to ours \citep{gusev_efremov_2013, gusev_2022}, where the spiral arm was parameterised in the $(\ln R, \theta)$ plane. It was found that adjacent star-forming regions in all four galaxies were either at a spacing of $350-500\,\si{pc}$ and/or integer multiples (2-4) of this range. This range of separations is remarkably similar to what we find, that is, $\simeq 500\, \si{pc}$ in the hydro and $\simeq 650 \,\si{pc}$ in the weakly magnetised run. This is in spite of the differences between the parameters of our models and these galaxies. 

We found that the Fourier transform of the column density along the spiral arms exhibits peaks that are integer multiples of each other (c.f. Fig.~\ref{fig:power_spectrum}). This effect has also been reported in all the four galaxies mentioned here \citep[see][]{gusev_2022}. Since this is seen in both the hydro and the weakly magnetised cases, we think magnetic fields are not the main cause behind this, as previously suggested in \cite{efremov_2010}. This intriguing trend is also reflected in the strings of $\rm H\textsc{i}$ super-clouds found in the Carina arm of the Milky Way, separated by $700~\si{pc} \pm 100~\si{pc}$, where more massive clouds are at about twice this separation \citep{geumsook_2023}. To draw firm conclusions from the cloud separation themselves, however, we need to expand the parameter space and tune our models to different nearby galaxies. This will help us understand the dependence of the cloud separation on the global properties of the galaxy. 

\subsubsection{Comparison to simulations}
Other numerical works have reported an increase in the cloud separation by a factor of 2 \cite[in local 2D simulations by][]{kim_wong_kim_2015} and 3 \cite[in local 3D simulations by][]{kim_ostriker_2006} in spiral arms in the presence of magnetic fields for plasma-beta values similar to ours. This is larger than what we find, i.e., a factor of $1.3$ increase in the adjacent cloud separation as well as in the average unstable mode along the spiral arms. This could be due to the limited number statistics of the clouds in their local simulation boxes ($\lesssim 10$) or other limitations introduced by the local approximation, compared to global disc simulations. 

\subsection{Caveats} \label{subsect:caveats}

Since we focus on the effects of magnetic fields on the formation of clouds, for the sake of simplicity and computational costs, we do not include a star formation model or various feedback mechanisms, such as supernova feedback, ionising radiation, winds from massive stars, or cosmic rays. These processes could potentially have an impact on the fragmentation process of the clouds, and the structure of the spriral arms themselves. Such models including star formation and feedback will be considered in future studies. Here we focus purely on the effects of gas dynamics, which suggests that at least the onset of fragmentation of the sprial arms can be explained by self-gravity, cooling, and magnetic fields alone.

Our galaxies are gravitationally stable, have low shear (compared to the Milky Way), and are in the thermally stable regime initially. This is done to ensure that our galaxy is dominated by the spiral arms since we focus on the spiral arm instability in a global setting. 

\section{Summary} \label{sec:conclusions}
We study isolated spiral disc galaxies in global three-dimensional simulations with self-gravity, magnetic fields, equilibrium heating and cooling, and an external spiral potential, to study the impact of varying magnetic field strengths on the spiral arm instability. The spiral arms in our simulations form self-consistently and fragment into beads-on-a-string patterns. We find that the magnetic fields have a major dynamical impact on the spiral arm instability, which mainly depends upon their initial strength. Our conclusions are summarised as follows:
\begin{itemize}
    \item For comparable spiral background potentials, we find that moderate initial magnetic fields ($\beta = 10$) stabilise the spiral arms against fragmentation, in contrast to the hydro and the weak-field case ($\beta = 50$), where the arms are unstable. The moderate magnetic field case forms arms that are more diffuse and hotter compared to the other cases, due to the additional opposing magnetic pressure.
    \item For the case of weak initial magnetic fields ($\beta = 50$), the spiral arms fragment in the presence of amplified equipartition magnetic fields in the arms ($\beta_{\rm arm} \sim 2.3$). The magnetic tension of the fields stabilises the vortical KHI-like wiggles present in the hydro case. 
    \item We estimate the adjacent cloud separations in the un-magnetised (hydro) case to be $\sim 500\,\si{pc}$, and $\sim 650\,\si{pc}$ in the weakly-magnetised case. This is remarkably close to the separations observed in many nearby spiral galaxies, which show separations of star-forming regions in the range \mbox{$400$--$700\,\si{pc}$}.
    \item We find that the wavelength of the average unstable mode along the spiral arms increases in length from $0.73^{+0.31}_{-0.36}\,\si{kpc}$ in the hydro case, to $0.98^{+0.49}_{-0.46}\,\si{kpc}$ in the weakly-magnetised case.
    \item Additionally, we find that the peaks of the 1D Fourier power spectrum of the column density along the spiral arms show peaks that are integer multiples of each other for both the magnetic and un-magnetised cases. This has been reported for the nearby galaxies analysed for the regularity of star-forming regions along their spiral arms. 
    \item The spiral arms in the weakly-magnetised case separate out into disjointed clouds along the arms around $\sim 20\,\si{Myr}$ before the hydro case. We find evidence that this may be due to the onset of the Parker instability in the spiral arms. The calculated linear growth rates and length scales of the Parker instability fall within the expected values seen in the simulation. We also show the magnetic field morphology around the clouds in the arm that form magnetic hills and valleys (c.f., Fig.~\ref{fig:parker_instability_3D}), as expected from linear theory. 
\end{itemize}

With the advent of the James Webb Space Telescope, we can now resolve infrared cores observed along the spiral arms of nearby galaxies \citep{elmegreen_2006, elmegreen_2018, elmegreen_2019} in unprecedented detail. Future parameter studies will aim to tailor our models to nearby galaxies for a more direct comparison with these observations.

\begin{acknowledgements}
C.~F.~acknowledges funding provided by the Australian Research Council (Future Fellowship FT180100495 and Discovery Project DP230102280), and the Australia-Germany Joint Research Cooperation Scheme (UA-DAAD). We further acknowledge high-performance computing resources provided by the Leibniz Rechenzentrum and the Gauss Centre for Supercomputing (grants~pr32lo, pr48pi and GCS Large-scale project~10391), the Australian National Computational Infrastructure (grant~ek9) and the Pawsey Supercomputing Centre (project~pawsey0810) in the framework of the National Computational Merit Allocation Scheme and the ANU Merit Allocation Scheme. The simulation software, \texttt{FLASH}, was in part developed by the Flash Centre for Computational Science at the Department of Physics and Astronomy of the University of Rochester.
\end{acknowledgements}

\bibliographystyle{aa}
\bibliography{feather_biblio,federrath}

\begin{thebibliography}{53}
\expandafter\ifx\csname natexlab\endcsname\relax\def\natexlab#1{#1}\fi

\bibitem[{{Beck}(2015)}]{beck_2015}
{Beck}, R. 2015, \aapr, 24, 4

\bibitem[{Beck {et~al.}(2020)Beck, Chamandy, Elson, \& Blackman}]{beck_chamandy_elson_2020}
Beck, R., Chamandy, L., Elson, E., \& Blackman, E.~G. 2020, Galaxies, 8

\bibitem[{{Beck} {et~al.}(1989){Beck}, {Loiseau}, {Hummel}, {Berkhuijsen}, {Grave}, \& {Wielebinski}}]{beck_1989}
{Beck}, R., {Loiseau}, N., {Hummel}, E., {et~al.} 1989, \aap, 222, 58

\bibitem[{{Binney} \& {Merrifield}(1998)}]{binney_1998}
{Binney}, J. \& {Merrifield}, M. 1998, {Galactic Astronomy}

\bibitem[{{Chakrabarti} {et~al.}(2003){Chakrabarti}, {Laughlin}, \& {Shu}}]{chakrabarti_2003}
{Chakrabarti}, S., {Laughlin}, G., \& {Shu}, F.~H. 2003, \apj, 596, 220

\bibitem[{{Cox} \& {G{\'o}mez}(2002)}]{cox_gomez_2002}
{Cox}, D.~P. \& {G{\'o}mez}, G.~C. 2002, \apjs, 142, 261

\bibitem[{{Dobbs} \& {Price}(2008)}]{dobbs_price_2008}
{Dobbs}, C.~L. \& {Price}, D.~J. 2008, \mnras, 383, 497

\bibitem[{{Dubey} {et~al.}(2008){Dubey}, {Fisher}, {Graziani}, {Jordan}, {Lamb}, {Reid}, {Rich}, {Sheeler}, {Townsley}, \& {Weide}}]{dubey_fisher_2008}
{Dubey}, A., {Fisher}, R., {Graziani}, C., {et~al.} 2008, in Astronomical Society of the Pacific Conference Series, Vol. 385, Numerical Modeling of Space Plasma Flows, ed. N.~V. {Pogorelov}, E.~{Audit}, \& G.~P. {Zank}, 145

\bibitem[{{Efremov}(2009)}]{efremov_2009}
{Efremov}, Y.~N. 2009, Astronomy Letters, 35, 507

\bibitem[{{Efremov}(2010)}]{efremov_2010}
{Efremov}, Y.~N. 2010, \mnras, 405, 1531

\bibitem[{{Elmegreen} \& {Elmegreen}(1983)}]{elmegreen_1983}
{Elmegreen}, B.~G. \& {Elmegreen}, D.~M. 1983, \mnras, 203, 31

\bibitem[{{Elmegreen} \& {Elmegreen}(2019)}]{elmegreen_2019}
{Elmegreen}, B.~G. \& {Elmegreen}, D.~M. 2019, \apjs, 245, 14

\bibitem[{Elmegreen {et~al.}(2018)Elmegreen, Elmegreen, \& Efremov}]{elmegreen_2018}
Elmegreen, B.~G., Elmegreen, D.~M., \& Efremov, Y.~N. 2018, The Astrophysical Journal, 863, 59

\bibitem[{Elmegreen {et~al.}(2006)Elmegreen, Elmegreen, Kaufman, Sheth, Struck, Thomasson, \& Brinks}]{elmegreen_2006}
Elmegreen, D.~M., Elmegreen, B.~G., Kaufman, M., {et~al.} 2006, The Astrophysical Journal, 642, 158

\bibitem[{{Federrath}(2016)}]{Federrath2016jpp}
{Federrath}, C. 2016, Journal of Plasma Physics, 82, 535820601

\bibitem[{{Federrath} {et~al.}(2011){Federrath}, {Chabrier}, {Schober}, {Banerjee}, {Klessen}, \& {Schleicher}}]{FederrathEtAl2011}
{Federrath}, C., {Chabrier}, G., {Schober}, J., {et~al.} 2011, \prl, 107, 114504

\bibitem[{{Federrath} {et~al.}(2010){Federrath}, {Roman-Duval}, {Klessen}, {Schmidt}, \& {Mac Low}}]{FederrathDuvalKlessenSchmidtMacLow2010}
{Federrath}, C., {Roman-Duval}, J., {Klessen}, R.~S., {Schmidt}, W., \& {Mac Low}, M. 2010, \aap, 512, A81

\bibitem[{{Federrath} {et~al.}(2022){Federrath}, {Roman-Duval}, {Klessen}, {Schmidt}, \& {Mac Low}}]{FederrathEtAl2022ascl}
{Federrath}, C., {Roman-Duval}, J., {Klessen}, R.~S., {Schmidt}, W., \& {Mac Low}, M.~M. 2022, {TG: Turbulence Generator}, Astrophysics Source Code Library, record ascl:2204.001

\bibitem[{{Fryxell} {et~al.}(2000){Fryxell}, {Olson}, {Ricker}, {Timmes}, {Zingale}, {Lamb}, {MacNeice}, {Rosner}, {Truran}, \& {Tufo}}]{FryxellEtAl2000}
{Fryxell}, B., {Olson}, K., {Ricker}, P., {et~al.} 2000, \apjs, 131, 273

\bibitem[{{Gusev} \& {Efremov}(2013)}]{gusev_efremov_2013}
{Gusev}, A.~S. \& {Efremov}, Y.~N. 2013, \mnras, 434, 313

\bibitem[{{Gusev} \& {Shimanovskaya}(2020)}]{gusev_shimanocskayay_2020}
{Gusev}, A.~S. \& {Shimanovskaya}, E.~V. 2020, \aap, 640, L7

\bibitem[{{Gusev} {et~al.}(2022){Gusev}, {Shimanovskaya}, \& {Zaitseva}}]{gusev_2022}
{Gusev}, A.~S., {Shimanovskaya}, E.~V., \& {Zaitseva}, N.~A. 2022, \mnras, 514, 3953

\bibitem[{{Hanawa} \& {Kikuchi}(2012)}]{hanawa_kikuchi_2012}
{Hanawa}, T. \& {Kikuchi}, D. 2012, in Astronomical Society of the Pacific Conference Series, Vol. 459, Numerical Modeling of Space Plasma Slows (ASTRONUM 2011), ed. N.~V. {Pogorelov}, J.~A. {Font}, E.~{Audit}, \& G.~P. {Zank}, 310

\bibitem[{{Khoperskov} \& {Khrapov}(2018)}]{khoperskov_2017}
{Khoperskov}, S.~A. \& {Khrapov}, S.~S. 2018, \aap, 609, A104

\bibitem[{{Kim} {et~al.}(2014){Kim}, {Kim}, \& {Kim}}]{kim_wong_kim_2014}
{Kim}, W.-T., {Kim}, Y., \& {Kim}, J.-G. 2014, \apj, 789, 68

\bibitem[{{Kim} \& {Ostriker}(2006)}]{kim_ostriker_2006}
{Kim}, W.-T. \& {Ostriker}, E.~C. 2006, \apj, 646, 213

\bibitem[{{Kim} {et~al.}(2015){Kim}, {Kim}, \& {Elmegreen}}]{kim_wong_kim_2015}
{Kim}, Y., {Kim}, W.-T., \& {Elmegreen}, B.~G. 2015, \apj, 809, 33

\bibitem[{{K{\"o}rtgen} {et~al.}(2019){K{\"o}rtgen}, {Banerjee}, {Pudritz}, \& {Schmidt}}]{bastian_2019}
{K{\"o}rtgen}, B., {Banerjee}, R., {Pudritz}, R.~E., \& {Schmidt}, W. 2019, \mnras, 489, 5004

\bibitem[{{Kosi{\'n}ski} \& {Hanasz}(2006)}]{hanasz_kosinki_2006}
{Kosi{\'n}ski}, R. \& {Hanasz}, M. 2006, \mnras, 368, 759

\bibitem[{{Koyama} \& {Inutsuka}(2002)}]{koyama_inutsuka_2002}
{Koyama}, H. \& {Inutsuka}, S.-i. 2002, \apjl, 564, L97

\bibitem[{Körtgen {et~al.}(2018)Körtgen, Banerjee, Pudritz, \& Schmidt}]{bastian_2018}
Körtgen, B., Banerjee, R., Pudritz, R.~E., \& Schmidt, W. 2018, Monthly Notices of the Royal Astronomical Society: Letters, 479, L40

\bibitem[{{Lee}(2014)}]{lee_kit_2014}
{Lee}, W.-K. 2014, \apj, 792, 122

\bibitem[{{Mandowara} {et~al.}(2021){Mandowara}, {Sormani}, {Sobacchi}, \& {Klessen}}]{mandowara_sormani_2021}
{Mandowara}, Y., {Sormani}, M.~C., {Sobacchi}, E., \& {Klessen}, R.~S. 2021, arXiv e-prints, arXiv:2110.04108

\bibitem[{Mouschovias(1996)}]{mouschovias1996}
Mouschovias, T.~C. 1996, The Parker Instability in the Interstellar Medium, ed. K.~C. Tsinganos (Dordrecht: Springer Netherlands), 475--504

\bibitem[{{Mouschovias} {et~al.}(2009){Mouschovias}, {Kunz}, \& {Christie}}]{mouschovias_2009}
{Mouschovias}, T.~C., {Kunz}, M.~W., \& {Christie}, D.~A. 2009, \mnras, 397, 14

\bibitem[{{Mulcahy} {et~al.}(2017){Mulcahy}, {Beck}, \& {Heald}}]{mulcahy_beck_2017}
{Mulcahy}, D.~D., {Beck}, R., \& {Heald}, G.~H. 2017, \aap, 600, A6

\bibitem[{Nguyen {et~al.}(2017)Nguyen, Pettitt, Tasker, \& Okamoto}]{ngan_pettitt_2017}
Nguyen, N.~K., Pettitt, A.~R., Tasker, E.~J., \& Okamoto, T. 2017, Monthly Notices of the Royal Astronomical Society, 475, 27

\bibitem[{{Park} {et~al.}(2023){Park}, {Koo}, {Kim}, \& {Elmegreen}}]{geumsook_2023}
{Park}, G., {Koo}, B.-C., {Kim}, K.-T., \& {Elmegreen}, B. 2023, arXiv e-prints, arXiv:2308.01577

\bibitem[{{Parker}(1966)}]{parker_1966}
{Parker}, E.~N. 1966, \apj, 145, 811

\bibitem[{{Parker}(1967{\natexlab{a}})}]{parker_1967_I}
{Parker}, E.~N. 1967{\natexlab{a}}, \apj, 149, 517

\bibitem[{{Parker}(1967{\natexlab{b}})}]{parker_1967_II}
{Parker}, E.~N. 1967{\natexlab{b}}, \apj, 149, 535

\bibitem[{{Parker}(1968{\natexlab{a}})}]{parker_1968_II}
{Parker}, E.~N. 1968{\natexlab{a}}, \apj, 154, 57

\bibitem[{{Parker}(1968{\natexlab{b}})}]{parker_1968_I}
{Parker}, E.~N. 1968{\natexlab{b}}, \apj, 154, 49

\bibitem[{{Proshina} {et~al.}(2022){Proshina}, {Moiseev}, \& {Sil'chenko}}]{proshina_moiseev_silchenko_2022}
{Proshina}, I.~S., {Moiseev}, A.~V., \& {Sil'chenko}, O.~K. 2022, Astronomy Letters, 48, 139

\bibitem[{{Renaud} {et~al.}(2013){Renaud}, {Bournaud}, {Emsellem}, {Elmegreen}, {Teyssier}, {Alves}, {Chapon}, {Combes}, {Dekel}, {Gabor}, {Hennebelle}, \& {Kraljic}}]{renaud_2013}
{Renaud}, F., {Bournaud}, F., {Emsellem}, E., {et~al.} 2013, \mnras, 436, 1836

\bibitem[{{Roberts}(1969)}]{roberst_69}
{Roberts}, W.~W. 1969, \apj, 158, 123

\bibitem[{{Shetty} \& {Ostriker}(2006)}]{shetty_ostriker_2006}
{Shetty}, R. \& {Ostriker}, E.~C. 2006, \apj, 647, 997

\bibitem[{Shukurov \& Subramanian(2021)}]{shukurov_subramanian_2021}
Shukurov, A. \& Subramanian, K. 2021, Astrophysical Magnetic Fields: From Galaxies to the Early Universe, Cambridge Astrophysics (Cambridge University Press)

\bibitem[{{Sormani} {et~al.}(2017){Sormani}, {Sobacchi}, {Shore}, {Tre{\ss}}, \& {Klessen}}]{sormani_sobacchi_2017}
{Sormani}, M.~C., {Sobacchi}, E., {Shore}, S.~N., {Tre{\ss}}, R.~G., \& {Klessen}, R.~S. 2017, \mnras, 471, 2932

\bibitem[{{Truelove} {et~al.}(1997){Truelove}, {Klein}, {McKee}, {Holliman}, {Howell}, \& {Greenough}}]{TrueloveEtAl1997}
{Truelove}, J.~K., {Klein}, R.~I., {McKee}, C.~F., {et~al.} 1997, \apjl, 489, L179

\bibitem[{{V{\'a}zquez-Semadeni} {et~al.}(2007){V{\'a}zquez-Semadeni}, {G{\'o}mez}, {Jappsen}, {Ballesteros-Paredes}, {Gonz{\'a}lez}, \& {Klessen}}]{VazquezSemadeniEtAl2007}
{V{\'a}zquez-Semadeni}, E., {G{\'o}mez}, G.~C., {Jappsen}, A.~K., {et~al.} 2007, \apj, 657, 870

\bibitem[{Wada(2008)}]{wada_2008}
Wada, K. 2008, The Astrophysical Journal, 675, 188

\bibitem[{{Wada} \& {Koda}(2004)}]{wada_koda_2004}
{Wada}, K. \& {Koda}, J. 2004, \mnras, 349, 270

\end{thebibliography}






   
  



\appendix

\section{Scale height estimation} \label{Appendix:scale_height_estimation}
\begin{figure}
    \centering
    \includegraphics[width = 0.95\linewidth]{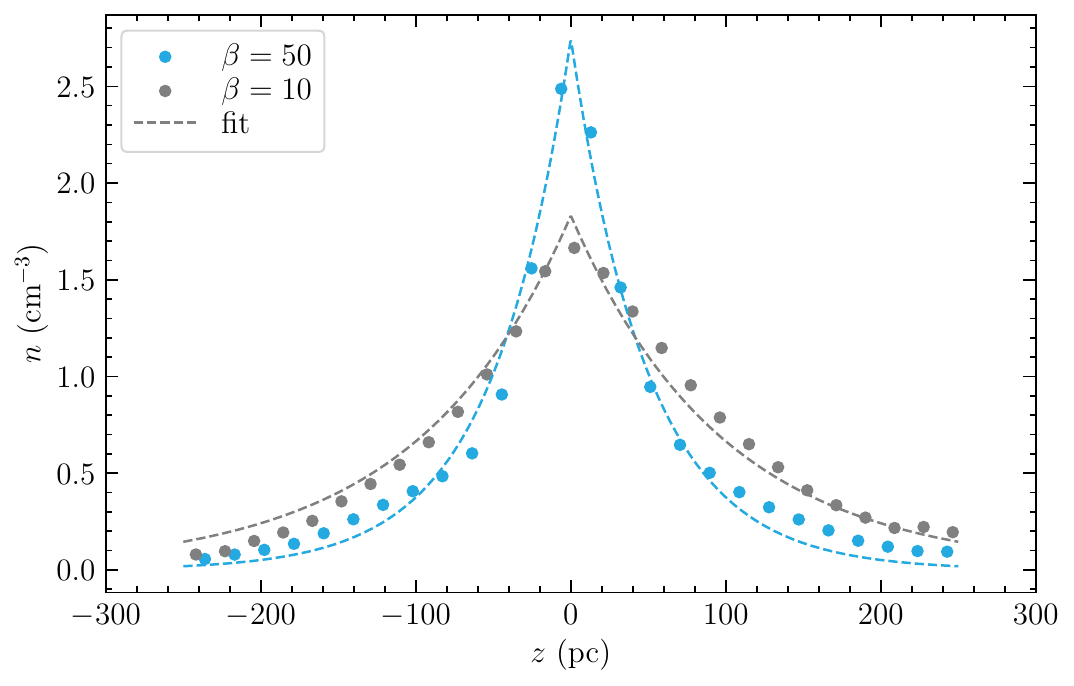}
    \caption{Gas density as a function of the z-coordinate around the traced spiral arms at $t = 0.50\,T_{\rm rot}$. The fit is of the functional form $\rho(z) = \rho_{0} \exp{(-|z|/H)}$. }
    \label{fig:scaleHeight_estimation}
\end{figure}

We estimate the scale height around the spiral arms by quantifying the density as a function of the z-coordinate, which is perpendicular to the plane of the galactic disc. We take the spiral arms traced via the friends of friends (FoF) algorithm (see \autoref{result_subsect_spiral_arm_properties}). For each (x,y) in the arms, we define $z = 0$ as the point of maximum density. This makes sure that we trace the spiral arms in three dimensions. We then bin the density in bins of $~20\,\rm pc$. This is shown in \autoref{fig:scaleHeight_estimation}, where the density is plotted on the vertical axis and the z-coordinate on the horizontal axis. For the scale height calculation we then fit the functional form $\rho(z) = \rho_{0} \exp{(-|z|/H)}$ to the binned data. The fits are shown as dotted lines in \autoref{fig:scaleHeight_estimation}. This gives us the $\rho_{0}$ and the scale height $H$. The scale heights are found to be $50.2 \pm 2.2$~pc for the $\beta = 50$ case and $98.6 \pm 3.9$~pc for the $\beta = 10$ case.

\end{document}